\documentclass{aa}  
\usepackage{graphicx}
\usepackage{txfonts}
\usepackage{hyperref}
\usepackage{natbib}  
\usepackage{paralist}

\newcommand{\textbfediting}{}%\textcolor{red}}
\newcommand\kepler{\textsl{Kepler}}

\usepackage{gensymb}

\usepackage{ulem}
\begin{document}

%%-----------------------------------------------------------------
%%      the top matter
%%
% \linenumbers
%   \title{The impact of magnetic fields on  mixed-mode frequencies}
\title{Magnetic signatures on mixed-mode frequencies}
   \subtitle{II. Period spacings as a probe of the internal magnetism of red giants }
%   \title{The impact of magnetic fields on dipolar mixed-mode frequencies}
%   \subtitle{I. An axisymmetric poloidal and toroidal fossil field inside the core of red giants.}
   \titlerunning{Magnetic signatures on mixed-mode frequencies II. Period spacings}
   
% \title{Towards an asteroseismic detection of magnetism in the core of red giants}
%   \subtitle{I. An axisymmetric fossil field inside the core of red giants.}
%   \title{The impact of magnetic fields on dipolar mixed-mode frequencies}
%   \subtitle{I. An axisymmetric poloidal and toroidal fossil field inside the core of red giants.}
%   \titlerunning{Towards an asteroseismic detection of magnetism in the core of red giants}

   \author{L.~Bugnet
        %   \inst{1}
          %\and ...
        %   \and V.~Prat\inst{1}
        %   \and S.~Mathis\inst{1}
        %   \and A.~Astoul\inst{1}
        %   \and K.~Augustson\inst{1}
        %   \and R.~A.~Garc\'\i a\inst{1}
        %   \and S.~Mathur\inst{3,}\inst{4}
        %   \and L.~Amard\inst{5}\inst{} 
        %   \and C.~Neiner\inst{6} 
        %   \and M. J. Thompson\inst{6}\inst{{\dagger}} 
          }

\institute{%AIM, CEA, CNRS, Université Paris-Saclay, Université Paris Diderot, Sorbonne Paris Cité, F-91191 Gif-sur-Yvette, France
{Flatiron Institute, Simons Foundation, 162 Fifth Ave, New York, NY 10010, USA},\\ \email{lbugnet@flatironinstitute.org}}
% \and
% Instituto de Astrof\'{\i}sica de Canarias, E-38200, La Laguna, Tenerife, Spain
% \and 
% Universidad de La Laguna, Dpto. de Astrof\'{\i}sica, E-38205, La Laguna, Tenerife, Spain
% \and University of Exeter, Department of Physics and Astronomy, Stoker Road, Devon, Exeter, EX4 4GL,  UK
% \and LESIA, Observatoire de Paris, Université PSL, CNRS, Sorbonne Université, Université de Paris, 5 place Jules Janssen, F-92195 Meudon, France}
% \and {$\dagger$} High Altitude Observatory, National Center for Atmospheric Research, P.O. Box 3000, Boulder, CO 80307}

% \and 
% Space Science Institute, 4750 Walnut Street Suite 205, Boulder, CO 80301, USA}
\authorrunning{L. Bugnet}
   \date{Received 21 January 2022 / Accepted 24 August 2022}

%\author{A. Author1}\address{Timberland Observatory, 34560 City, Neverland}

%\author{J.-P. Author2}\address{Institute XYZ, 1299 City, OtherLand}

%% IF Author3 has the same affiliation than Author1:
%\author{C.\,E. Author3$^1$}

%% IF Author3 has its own affiliation:
%\author{C.\,E. Author3}\address{Dept. of Chess, University of Games, 35101 Las Vegas, Monaco} 

%% IF Author3 has two affiliations, the one of Author1 and a second one:
%\author{C.\,E. Author3$^{1,}$}\address{Dept. of Chess, University of Games, 35101 Las Vegas, Monaco} 

%% Keep this line, even if the page will be settled afterwards.
% \setcounter{page}{237}

%%-----------------------------------------------------------------

%%-----------------------------------------------------------------
%%        The abstract
%% 
%%  Warning!  within the abstract:
%%  - do not use macros. 
%%  - do not use commands like: \cite, \citet, \citep ... etc.

\abstract{Theoretical works have looked into the various topologies and amplitudes, as well as the stability of the magnetic field that is expected to be present in the radiative interior of stars evolving after the main sequence. From these studies, we know that strong stable "fossil" fields might be trapped inside evolved stars. These could trigger the strong transport of angular momentum from the core to the envelope, a process that is not generally included in state-of-the-art stellar models. This may therefore have a substantial impact on the mixing and the inferred stellar parameters. Such internal magnetic fields have never been observed in evolved stars. As a result, there is a major  piece missing from our global picture of stars as dynamical bodies.}
{Asteroseismology has opened a window onto stellar internal dynamics, as oscillation frequencies, amplitudes, and lifetimes are affected by processes that are taking place inside the star. The detection of buried magnetic fields could therefore be possible through the measurement of their impact on the oscillations of stars. This advancement would be groundbreaking for our knowledge of stellar dynamics. In this scope, magnetic signatures on mixed-mode frequencies have recently been characterized, but the task of detection  remains challenging as the mixed-mode frequency pattern is highly complex and affected by rotational effects, while modes of different radial orders are often intertwined. In this work, we aim to build a bridge between theoretical prescriptions and complex asteroseismic data analysis to facilitate a future search and characterization of internal magnetism with asteroseismology. }
{We investigated the effect of magnetic fields inside evolved stars with solar-like oscillations on the estimation of the period spacing of gravity-mode (g-mode) components of simulated mixed gravito-acoustic modes. We derived a new corrected stretching function of the power spectrum density to account for the presence of magnetic signatures on their frequencies.}
{We demonstrate that the strong dependency of the amplitude of the magnetic signature with mixed-mode frequencies leads to biased estimates of period spacings towards lower values. %, with the shift proportional to the square of the magnetic field amplitude and to the typical oscillation frequency following the $\nu^{-3}$ law. 
We also show that a careful analysis of the oscillation frequency pattern through various period spacing estimates and across a broad frequency range might lead to the first detection of magnetic fields inside red giants and at the same time, we adjust the measured value of g-mode period spacing. }
{}

%Most of the few hundred millions NASA TESS mission targets have never been specifically pointed by any telescope before. Hence, the amount of data to be analyzed is about to increase by several order of magnitude, compared to the few hundred thousand stars of CoRoT, \emph{Kepler} and K2 missions. It is of great matter to ease stars analysis by determining the nature of each target in nearly real time. We propose an automatic method to point out Solar-like pulsators among TESS targets. It relies on the use of the gloabl amount of power contained in power spectrum (already known as the FliPer method) as parameter fed into a machine learning classifier. Our study, based on TESS simulated datasets, show that we are able to distinguish pulsators with a $97\%$ accuracy.}

%% Insert the keywords (to appear in the ADS indexing)
%% Keywords must be separated by a comma

  \maketitle

\section{Introduction}
%  A RELIRE,  PREMIERE VERSION DE L'INTRODUCTION
% DRAFT:
% Stellar AM transport important for ages
% To probe interior properties of stars: mixed modes on the RGB
% Allows measure of rotation

% Recently, theoretical effect of axisymmetric magnetic field
% No detection yet

% e investigate effect of combined rotation and magentism on the observed period spacing of mixed modes
% Propose a method to detect magnetic sigantures on mm frequencies.

Unveiling the internal dynamical processes taking place within stellar interiors is one of the challenges related to the revolution in asteroseismology revolution over the past ten years \citep{Aerts2019}. It is of paramount importance
to achieve realistic modelings of the internal transport of angular momentum (AM) and mixing of chemical elements to compute robust evolutionary tracks \citep{Eggenberger2015}, as well as to estimate stellar ages \citep{Lebreton2014} in an accurate way. The \textsl{Kepler} mission \citep{Borucki2008} carried out observations of stars along their evolutionary tracks over four continuous years. This observing duration is large enough to allow the detection of very detailed patterns in the frequency domain, where stellar oscillations are studied. Asteroseismology recently allowed the measurement of the internal rotation rate of subgiants \citep[SG, e.g.,][]{Deheuvels2012a, Deheuvels2014a, Deheuvels2016, Deheuvels2020} and red giants (RG, e.g., \citealp{Beck2012, Beck2014, Mosser2012b, Mosser2015, Mosser2018,  Vrard2016, DiMauro2016b, Triana2017,  Gehan2018, Tayar2019}; see \citealp{Aerts2019} for a summary). These rotation-rate measurements point out a major discrepancy, namely, of two orders of magnitude, between the observed rotation rates and the much larger ones obtained through state-of-the-art stellar modeling \citep{Ceillier2012, Marques2013a, Eggenberger2012, Eggenberger2019a, Eggenberger2019b}. Therefore, there must be at least one process that would be efficient enough to transport angular momentum from the deepest layers of the star to the surface, which has not yet been taken into account in our current knowledge of stellar dynamics. This is despite the fact that numerous transport processes have been investigated to try and fill in the picture of stellar dynamics \citep{Mathis2013b, Aerts2019}:\\

% \begin{itemize}
 First, stellar model physics already includes rotation and associated hydrodynamical transport processes \citep[meridional circulation or shear instabilities, e.g.,][]{Zahn1992, Maeder1998, Mathis2004, Mathis2004b, Mathis2018a}, whose present characterizations do not include the observed low internal rotation rates of stellar cores or  weak core-to-surface differential rotation among their effects. The shear \citep{Zahn1992, Maeder1996, Talon1997, Mathis2004b, Mathis2018a} or the Goldreich-Schubert-Fricke \citep{Barker2019} instabilities are known to transport AM, but the expected effects are too weak for either of them to fully explain the observed rotation rates of solar-like stars \citep{Marques2013a}.

Second, mixed acoustic-gravity oscillation modes \citep{Belkacem2015} efficiently transport angular momentum during the RG phase, but their effect is too weak on the SG phase, where strong angular momentum transport is expected. On the contrary, internal gravity waves \citep{Talon2008, Pincon2017} are efficient in the early stages of evolution, but fail to transport AM during the red giant branch \citep[RGB,][]{Fuller2014}. 

Next, \cite{Tayler1973} suggested that a magneto-hydrodynamic instability might occur in radiative zones and \cite{Spruit2002} has demonstrated that the instability might cause a "Tayler-Spruit" dynamo. The physical existence of such magnetic dynamos has been debated by \cite{Zahn2007}, leading \cite{Fuller2019a} and \cite{Eggenberger2020} to investigate the effect of the instability itself (without the dynamo loop) on the AM transport inside RGs. The efficiency of the Maxwell stresses on transporting AM is in good agreement with observations, but this instability cannot  simultaneously solve the AM transport problem in SGs and RGs \citep{Eggenberger2019b, Eggenberger2020}. 
 
 Finally, stable magnetic fields tend to homogenize rotation along the field lines \citep{Mestel1987, Gough1990a, Brun2006a, Strugarek2011, Acevedo-Arreguin2013, Kissin2015a}, which could result in a highly efficient AM transport. Such stable fossil fields are observed at the surface of $\sim 10\%$ of intermediate-mass stars on the main sequence \citep[MS, e.g.,][]{Donati2009}. Inside RGs, magnetic fields could result from past dynamo action during the pre-main sequence (PMS) or MS and remain trapped inside the radiative interior of solar-like stars for the rest of their evolution (\citealp{MacGregor2002}; see the discussion on the magnetic scenario in \citealp{Bugnet2021}).
% \citep[see the discussion about magnetic scenario in][]{Bugnet2021}.
% \end{itemize}

In this work, we proceed under the hypothesis that stars possess magnetized stellar radiative interiors during the advanced stages of their evolution, namely, after the MS. We focus on red giant stars that present "solar-like oscillations" with visible mixed gravito-acoustic dipolar modes. This is the case for stars on the SG phase on the RGB, and the first and second clumps. The only window we might have on internal magnetism is magneto-asteroseismology \citep{Neiner2015,Mathis2021}, which consists of a search for characteristic signatures of magnetic fields in the observed frequency spectra of stellar oscillations. So far, only a few magneto-asteroseismology analyses have been performed, namely, for the Sun \citep{Goode1992,Kiefer2018} and for early-type stars \citep{Takata1994,Shibahashi2000}. Recent theoretical works on evolved stars \citep[red giants,][]{Bugnet2021, Mathis2021, Loi2020, Loi2021} have paved the way for observational studies of magnetism inside evolved stars. %We investigate the detectability of buried fields with magneto-asteroseismology. 
\cite{Fuller2015} \citep[and later][]{Loi2017, Lecoanet2017b} demonstrated that low-frequency internal gravity waves might be converted into Alfvén waves in the presence of strong magnetic fields. As a result, classical mixed acoustic gravity modes \sout{\textbfediting{(g-modes)}} might not form inside magnetized red giants. This process could impact the amplitudes of the observed dipolar and quadripolar mixed modes and may then justify the large fraction of observed intermediate-mass stars on the RGB showing abnormally low non-radial oscillation amplitudes \citep{Mosser2012, Garcia2014b, Stello2016a, Stello2016d}. To prove or disprove the hypothesis of magnetized radiative interiors and the resulting observations, recent studies have focused on the theoretical impact of stable fossil fields on mixed-mode frequencies.
Among them, \cite{Bugnet2021} demonstrated that the effect of the expected stable magnetic configuration (axisymmetric, aligned with the rotation axis of the star) on mixed-mode frequencies is large enough to allow detections in \textsl{Kepler} data. \cite{Mathis2021} provided the theoretical framework needed to perform radial magnetic-field inversions from observed magnetic frequency perturbations. This prescription would be game-changing for the characterization of magnetic fields in the case of detections of frequency perturbations in asteroseismic data. \cite{Loi2021} completed the theoretical picture of the effect of fossil fields on the mixed-mode frequency pattern by considering the effect of the field inclination on the magnetic signature, similarly to the study of \cite{Goode1992}.

Despite all this recent progress in terms of theoretical study, the internal magnetic fields of RGs have not yet been detected. Such detections require a very  detailed analysis of individual mode frequencies, which is even more challenging than recent studies that have allowed  for the measurement of RG core rotation rates \citep{Vrard2016, Mosser2018, Gehan2018}. Thus, the search for internal magnetism requires more guidance. This study is aimed at building a bridge between theoretical predictions and the complexity and challenges related to asteroseismic observables.

Mixed modes result from a coupling between acoustic modes \textbfediting{(p-modes)} that are evenly spaced in frequency and \textbfediting{gravity modes (g-modes)} in the radiative interior that are evenly spaced in terms of period \citep[e.g.,][]{Dupret2009}. A characteristic frequency spacing ($\Delta\nu$) is associated with the acoustic pattern and a characteristic period spacing \citep[defined as the distance in period between two consecutive g-modes with the same horizontal order and consecutive radial orders $\Delta\Pi_1$, ][]{Beck2011} is associated with pure g-modes. The g-mode period spacing can be measured from asteroseimic observation \citep[e.g.,][]{Mosser2015, Vrard2016, Buysschaert2016} and its value reflects the inner structure of the star along its evolution \citep{Bedding2011, Mosser2012c, Stello2013a}.\\

In Section~\ref{sec:periodspacingtheory}, we provide the physical background required to generate artificial RG frequency patterns on which we base this work. In Section~\ref{sec:mageffecttheory}, we use this typical RG frequency pattern to investigate the effect of a fossil stable magnetic field in the radiative interior on the determination of the g-mode period spacing. In Section~\ref{sec:probeB}, we demonstrate that magnetic signatures could possibly be detected thanks to the effect they induce on the observed g-mode period spacing. We discuss results in Sect.~\ref{sec:result}. In Sect.~\ref{sec:ccl}, we  conclude with an observational perspective on this work.
\section{Measurement of the g-mode period spacing}
\label{sec:periodspacingtheory}
%State-of-the-art $\Delta\Pi_1$ estimates on the RGB and associated confusions}
% \subsection{Background}
The objective of this study is to investigate the detectability of an axisymmetric magnetic field inside the radiative interior of evolved stars with solar-like oscillations from the measurement of the period spacing between pure g-modes (hereafter noted $\Delta\Pi_1$). % the way conducted by \cite{Mosser2015, Mosser2018, Vrard2016}. 
To remain unbiaised from observational interpretations, we base our study on simulated power spectral densities (PSD) of RGB stars through a magnetic adaptation of the \texttt{sloscillations}\footnote{\url{https://github.com/jsk389/sloscillations}} pipeline \citep{Kuszlewicz2018} thanks to the prescriptions of \cite{Bugnet2021} and \cite{Mathis2021}.

\subsection{Building synthetic asteroseismic power spectrum densities}
In this section, we provide the physical context and equations needed to build a realistic red-giant synthetic PSD and we rely on existing methods to extract the stellar seismic parameters of interest, $\Delta\Pi_1$, from an asteroseismic analysis.

\subsubsection{Mixed-mode frequencies}
\label{sec:mmf}
The second-order asymptotic expression for pure \textbfediting{p-modes} allows for an estimation of the frequencies of radial ($\ell=0$), dipolar ($\ell=1$), and quadripolar ($\ell=2$) modes following \cite{Mosser2011}. The asymptotic expression for dipolar-mode eigenfrequencies reads \citep{Tassoul1980}:
\begin{equation}
    \nu_{{\rm p}, \ell=1} = \left(n_{\rm p}+\frac{1}{2}+\epsilon_{\rm p} -d_{01}+\frac{\alpha}{2}\left(n_{\rm p}-n_{\rm max}\right)^2\right)\Delta\nu\, ,
    \end{equation}
    
    \noindent with $\nu_{{\rm p},\ell=1}$ as the theoretical frequency of pure pressure modes, $\Delta\nu$ as the frequency separation due to pressure modes of consecutive radial orders, $n_{\rm p}$ as the radial order of the pure pressure mode, $\epsilon_{\rm p}$ as the acoustic offset, $n_{\rm max}=\frac{\displaystyle \nu_{\rm max}}{\Delta\nu}-\epsilon_{\rm p}$ as the order corresponding to the p-$m$ mode closest to $\nu_{\rm max}$, $\alpha=0.015\Delta\nu^{-0.32}$ \citep{Mosser2013, Vrard2016}, and $d_{01}=0.0553-0.036\log{\Delta\nu,}$ as measured from stars on the RGB by \cite{Mosser2014, Mosser2018}.\\
    
    The structure of stars evolving as giants after the end of the main sequence allows interval gravity waves, propagating in the radiative core, to couple with the acoustic waves to form stationary mixed modes. Their theoretical frequencies are estimated following \cite{Mosser2012b} for dipolar so-called "mixed" modes as
\begin{equation}
\nu_{\rm{pg}, \ell=1}=\nu_{{\rm p},\ell=1} +\frac{\Delta\nu}{\pi} \arctan{\left[q\tan{\left[\pi\left(\frac{1}{\Delta\Pi_1 \nu_{\rm{pg}, \ell=1}}-\epsilon_{\rm g}\right)\right]}\right]}\, ,
\label{eq:nupg}
\end{equation}

\noindent where $\epsilon_{\rm g}$ is the gravity offset and $q$ is the coupling factor of mixed modes \citep{Mosser2017b} estimated by combining the phases $\theta_{\rm p}$ and $\theta_{\rm g}$ of pressure- and gravity-wave contributions to mixed modes, following:
\begin{equation}
    \tan{\theta_{\rm p}}=q\tan{\theta_{\rm g}}\, ,
\end{equation}\noindent with
\begin{equation}
    \theta_{\rm p} = \pi\frac{\nu_{\rm{pg}, \ell=1}-\nu_{{\rm p},\ell=1}}{\Delta\nu}
,\end{equation}
\noindent and
\begin{equation}
    \theta_{\rm g} = \pi\frac{1}{\Delta\Pi_1}\left(\frac{1}{\nu_{\rm{pg}, \ell=1}}-\frac{1}{\nu_{{\rm g},\ell=1}}\right)\, .
\end{equation}

In this study, we focus on dipolar mixed modes as they are the most sensitive to internal processes. We use the following notations: $\nu$ to simplify the writing of the dipolar mixed-mode frequencies ($\nu_{\rm{pg}, \ell=1}$), $\nu_{\rm p}$ for the frequency of pure $\ell=1$ pressure modes ($\nu_{{\rm p},\ell=1}$), and $\nu_{\rm g}$ for the frequency of pure $\ell=1$ g-modes ($\nu_{{\rm g},\ell=1}$).\\

{Using the elements introduced in Sections~\ref{sec:mmf} and Appendix~\ref{App:linewidthheight}, we can produce a synthetic oscillation-mode PSD of a typical RGB star with $\nu_{\rm max}=108.3\, \mu{\rm Hz}$, $\Delta\nu=10.84\, \mu$Hz, and $\Delta\Pi_1=75.2$ seconds. The result is represented on the top panel of Fig.~\ref{fig:PSDs} by considering the typical observable frequency range [~$\nu_{\rm max}-3\Delta\nu$~:~$\nu_{\rm max}+3\Delta\nu$~].}

\subsubsection{Mixed mode asymptotic behavior}
\label{sec:zeta}
Mixed-mode inertia can be dominated either by their acoustic component or by their gravity component, depending on their eigenfrequency. The nature of a mode can be measured from the comparison of the mode inertia \citep{Goupil2013a} in the radiative interior (mostly probed by g-modes) and the mode inertia in the convective envelope (mostly probed by \textbfediting{p-modes}). The ratio of the mode inertia in the radiative cavity on the total mode inertia defines the coupling function of mixed modes, expressed as \citep{Goupil2013a, Hekker2017}:
\begin{equation}
    \zeta = \left(1+\frac{q}{\mathcal N}\frac{1}{q^2\cos{\theta_{\rm p}}^2+\sin{\theta_{\rm p}}^2}\right)^{-1}\, .
\end{equation}

\begin{figure*}
    \centering
    \includegraphics[width=0.86\textwidth]{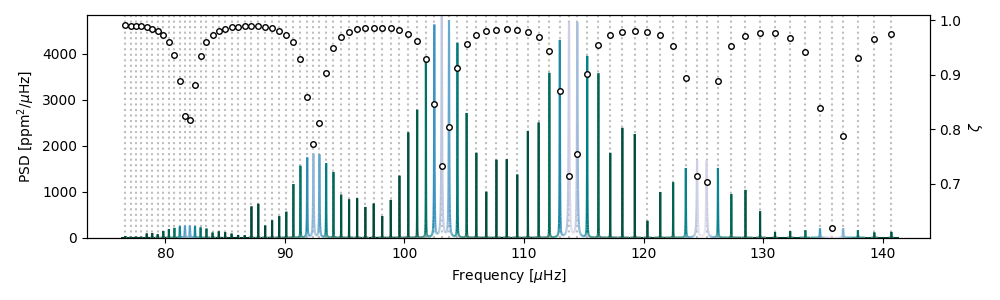}
    \includegraphics[width=0.86\textwidth]{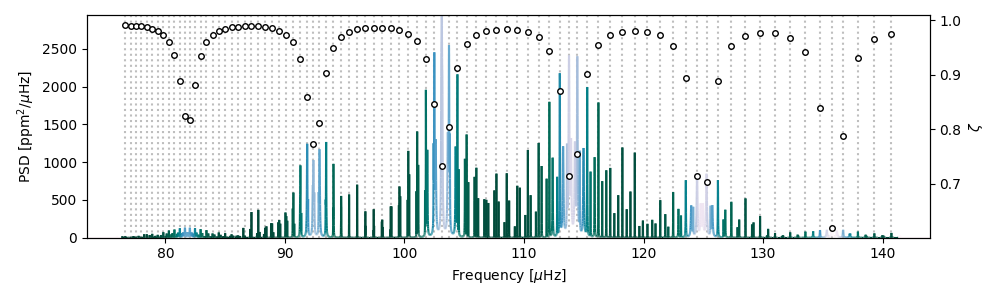}
    \includegraphics[width=0.86\textwidth]{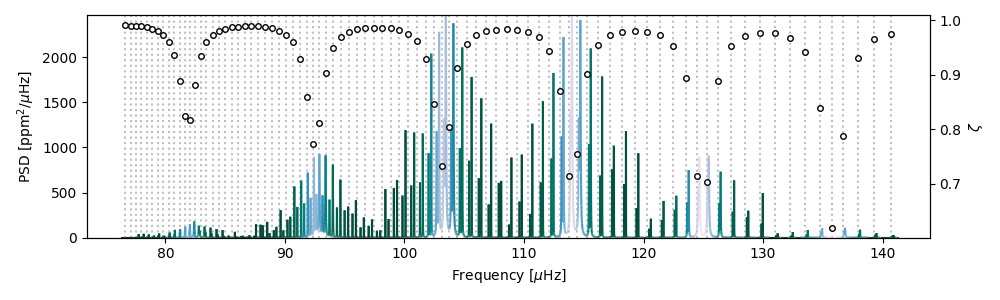}
    \includegraphics[width=0.86\textwidth]{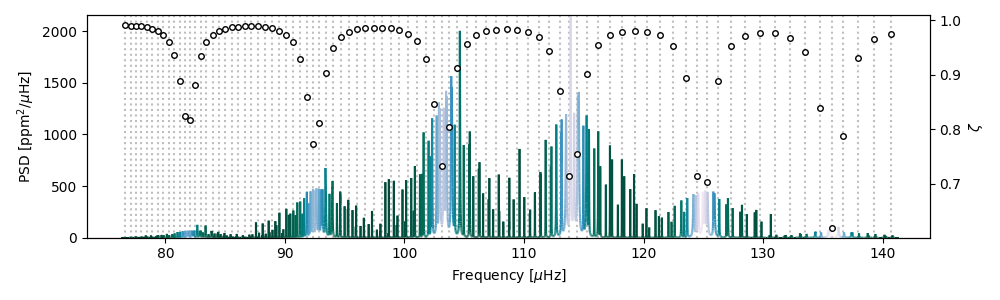}
    \caption{PSDs comprised of mixed dipolar oscillation modes of a simulated RGB star with $\nu_{\rm max}=110\,\mu$Hz and $\Delta\Pi_1=75.2$ sec. \textsl{Top panel:}   Non-rotating, non-magnetized model. \textsl{Second panel:} Model rotating with $\delta\nu_{\rm rot}=0.69\,\mu$Hz, $i=45\degree$. \textsl{Third panel:} Model agnetized with $\delta\nu_{\rm mag}(\nu_{\rm max})=50 \delta f_{\rm res}=0.4\,\mu$Hz, $i=45\degree$. \textsl{Bottom panel:} Model rotating and magnetized with $\delta\nu_{\rm rot}=0.69\,\mu$Hz, $\delta\nu_{\rm mag}(\nu_{\rm max})=0.4\,\mu$Hz, $i=45\degree$. The $\zeta$ function is overplotted on each panel to indicate the location of p-$m$ modes (dips) and g-$m$ modes (maxima), while the color of the PSD varies according to the zeta function value for better visualization of the nature of the modes (dark green: g-$m$ dipolar modes, light blue: p-$m$ dipolar modes). Dotted vertical lines indicate the location of the modes for the non-rotating, non-magnetized model.}
    \label{fig:PSDs}
\end{figure*}

The value of the $\zeta$ function is represented for each mixed mode of a simulated star on the RGB by the dots on each panel of Fig.~\ref{fig:PSDs}. When the mode is purely acoustic ($\theta_{\rm p}=0$), the $\zeta$ function can be rewritten as:
\begin{equation}
    \zeta_{\rm p} = \zeta_{\rm min} =\left(1+\frac{ \nu^2\Delta\Pi_1}{q\Delta\nu}\right)^{-1}\, .
    \label{eq:zeta}
\end{equation}
When the mode is a pure g-mode, $\left(\theta_{\rm p}=\displaystyle\frac{\pi}{2}\right)$, we have:\ 
\begin{equation}
    \zeta_{\rm g} = \zeta_{\rm max} = \left(1+\frac{q \nu^2\Delta\Pi_1}{\Delta\nu}\right)^{-1}\, .
\end{equation}
    The maxima of the $\zeta$ function therefore correspond to gravity-dominated (hereafter, g-$m$) modes, while dips are associated with acoustic-dominated (hereafter, p-$m$) modes. The g-$m$ modes are much more sensitive to internal processes than p-$m$ due to their relative inertia (Eq.~\ref{eq:zeta}). It is therefore convenient to focus on g-$m$ modes for the measurements of internal rotation rates, as done by \cite{Gehan2018}. We applied the same methodology to prevent surface dynamical processes from biasing the search for buried magnetic fields and we focused the rest of our study on the g-$m$ modes localized near maxima of the zeta function. We then suppressed all modes contained in the range [$\nu_{{\rm p}_{n}}-0.095\Delta\nu$, $\nu_{{\rm p}_{n}}+0.095\Delta\nu$] around the nominal $\nu_{{\rm p}_{n}}$ frequency at each radial order $n$ from the PSD before further analysis, as done by \cite{Gehan2018}. Radial and quadripolar modes are located in frequency ranges corresponding to g-$m$ dipolar modes. Thus, we also had to suppress their regions of the PSD for the measurement of $\Delta\Pi_1$ to be accurate. For each radial order, $n,$  we removed all the power in the frequency range of
[$\nu_{n,\ell=0}-0.2\Delta\nu$, $\nu_{n,\ell=0}+0.06\Delta\nu$], as done in \cite{Vrard2016}. The PSDs used in the following are thus solely composed of g-$m$ dipolar modes away from radial and quadripolar modes.\\
\subsubsection{Stretched period spacing of mixed modes}\label{sec:stretching}
While pure g-modes are evenly spaced in period, following: \begin{equation}
    \frac{1}{\nu_{\rm g}} = \left(-n_{\rm g}+\epsilon_{\rm g}\right)\Delta\Pi_1\, ,
\end{equation}
the g-mode period spacing is affected by the coupling with p-modes, as shown by Eq.~\ref{eq:nupg}. We can deduce the period of mixed modes $P$ as \citep{Mosser2015}:
\begin{equation}
    \frac{{\rm d}P}{{\rm d}n} = -\zeta \Delta\Pi_1\, ,
    \label{eq:dP}
\end{equation}
\noindent with $n=n_{\rm p}+n_{\rm g}$ the order of mixed modes.

Equation~\ref{eq:dP} enhances the observation that, due to the coupling of the modes, mixed modes are not evenly spaced in period. To extract the periodicity of mixed modes, \cite{Vrard2016}, \cite{Mosser2015}, and \cite{Hekker2017} therefore defined the "stretched" period $\pi$ via:
\begin{equation}
    {\rm d}\pi = -\frac{1}{\zeta}\frac{{\rm d}\nu}{\nu^2}\, .
    \label{eq:tau}
\end{equation}

\noindent The pure g-modes period spacing is expressed as \citep{Mosser2018}:
\begin{equation}
    \Delta\Pi_1 = -\int_{\displaystyle \nu_n}^{\displaystyle\nu_{n+1}}\frac{1}{\zeta}\frac{{\rm d}\nu}{\nu^2}.
    \label{eq:dpi}
\end{equation}
% We consider the variations of the $\zeta$ function near g-$m$ mode frequencies on consecutive orders to be small enough so that 
% \begin{equation}
%     \Delta\Pi_1 \approx -\frac{1}{\zeta(\nu_n)}\int_{\displaystyle \nu_n}^{\displaystyle\nu_{n+1}}\frac{{\rm d}\nu}{\nu^2}.
%     \label{eq:dpi_simp}
% \end{equation}
% \cite{Mosser2018} show the impact of such an assumption, >>>>> TO COMPLETE
% In the rest of our study, we use Eq.~\ref{eq:dpi_simp} instead of Eq.~\ref{dpi}, while keeping in mind the approximation.

\subsubsection{Measurement of $\Delta\Pi_1$ and associated uncertainty}
\label{sec:measure_dpi}

\cite{Vrard2016} measured the period spacing of g-modes by computing a Lomb-Scargle periodogram \citep[LS periodogram,][]{Lomb1976, Scargle1982} of the stretched PSD. Indeed, by stretching the original non-rotating, non-magnetized PSD onto the $\pi$ variable defined by Eq.~\ref{eq:tau}, the effect of the coupling of the modes on the observed frequencies through the $\zeta$ function is removed \citep{Vrard2016}, and the periodicity of g-modes is unveiled in the $\pi$-stretched PSD. The periodicity, $\Delta\Pi_1$, is then measured from the period associated with the maximum power in the LS periodogram.

The measurement of $\Delta\Pi_1$ from the stretching process is associated with various sources of uncertainties. We refer to Appendix A in the study of \cite{Vrard2016} for detailed discussions of all uncertainty sources in the measurement of this parameter from the LS method presented here. The resolution in the LS periodogram results is (in most cases) the main uncertainty contribution. Thus, this value has been used in most prior studies \citep{Vrard2016, Gehan2018}. The typical frequency of observed modes is $\nu_{\rm max}$, therefore, the LS periodogram has a resolution of $\nu_{\rm max}$ around $\Delta\Pi_1$. The nominal resolution expressed in the period variable is written as \citep{Vrard2016}:
\begin{equation}
    \delta\left(\Delta\Pi_1\right)_{\rm res}=\nu_{\rm max}\Delta\Pi_1^2\, .
    \label{eq:uncertainty_res}
\end{equation}

% We use this prescription to compute uncertainties on the measure of $\Delta\Pi_{1_{\rm ls}}$. 

The typical resolution uncertainty is about $0.5$ seconds on the RGB, representing a relative error of $1\%$ in the measurement of $\Delta\Pi_1$. While this seems reasonable, an error in the estimate of $\Delta\Pi_1$ of that order of magnitude would highly bias any search for signatures of internal dynamical processes. 
We performed an oversampling of the LS periodogram (by a factor of $6)$ to produce all results and figures in this article. This allowed us to get a more precise estimate of the location of the peak of maximum power, as demonstrated by \cite{Vrard2016}. Following \cite{Mosser2009b}, we then use the oversampling resolution:

\begin{equation}
    \delta\left(\Delta\Pi_{1_{\rm over}}\right) = \frac{1.6}{A}\delta\left(\Delta\Pi_{1_{\rm res}}\right)\, .
    \label{eq:uncertainty_over}
\end{equation}
{As discussed in \cite{Vrard2016}, typical dominant peak amplitudes $A$ are largely above the threshold justifying much {tighter resolution than the nominal resolution} in the LS periodogram given by Eq.~\ref{eq:uncertainty_res}.} 
% In the following sections~\ref{sec:rot} and \ref{sec:mag} we demonstrate the sensitivity of internal rotation and internal magnetism measurements to this precise estimate of $\Delta\Pi_1$.

% \subsubsection{Measure of the dipolar gravity-mode periodicity $\Delta\Pi_1$}
% % \subsubsection{Detection of the mixed mode periodocity}
% \label{sec:measure_dpi}

\subsection{Impact of the rotation on $\Delta\Pi_1$ measurements}
\label{sec:rot}

\begin{figure}[t]
    \centering
    \includegraphics[width=0.45\textwidth]{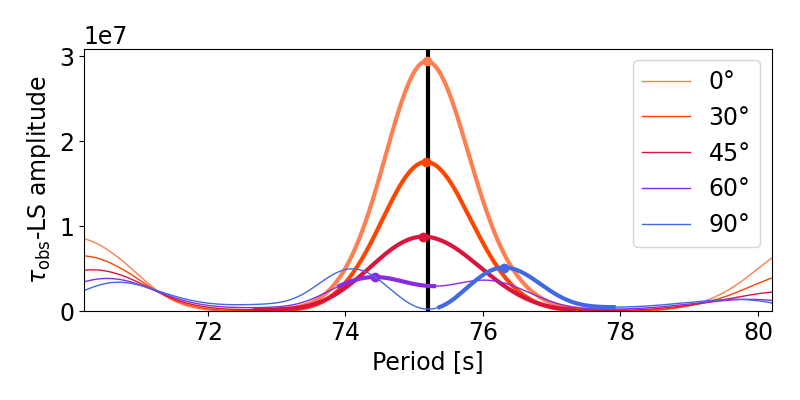}
    \includegraphics[width=0.45\textwidth]{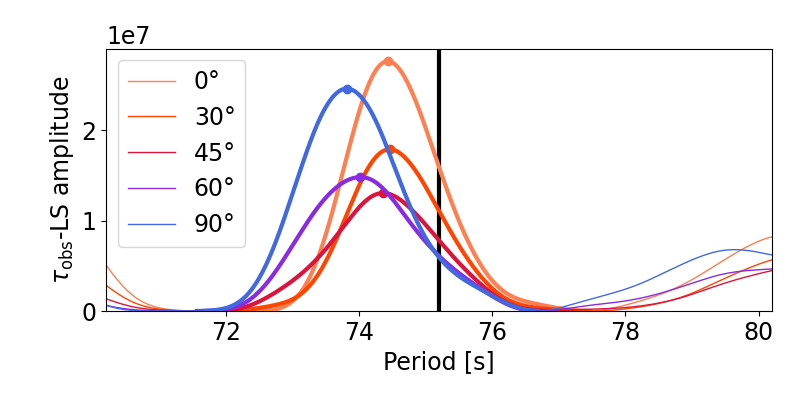}
    \includegraphics[width=0.45\textwidth]{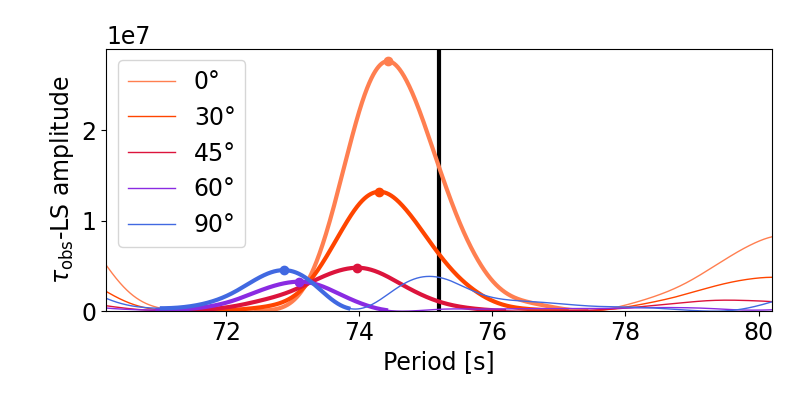}
    \includegraphics[width=0.45\textwidth]{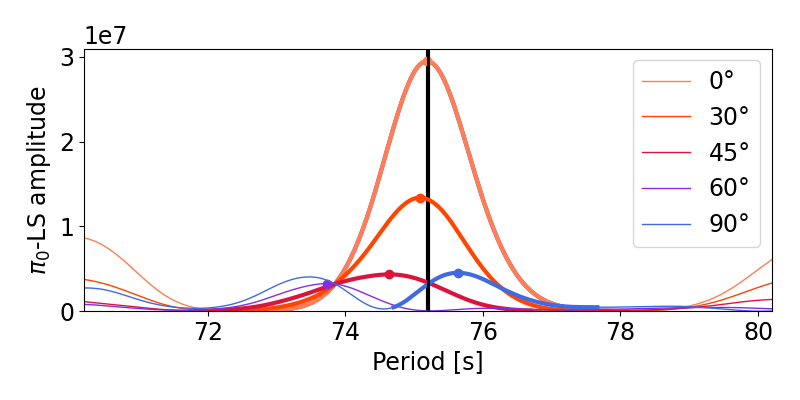}
    \caption{\textbfediting{LS periodograms for different stellar models.} \textsl{Top panel:} ${\rm d}\tau_{\rm obs}$-LS periodogram for a simulated star rotating such as $\delta\nu_{\rm rot,g}=0.69\, \mu$Hz. \textsl{Second panel:} Same diagram for a simulated magnetized, non-rotating star such as $\delta\nu_{\rm mag,g}(\nu_{\rm max})=0.4\, \mu$Hz. \textsl{Third panel:} Same diagram for a simulated magnetized and rotating star such as $\delta\nu_{\rm rot,g}=0.69\, \mu$Hz and $\delta\nu_{\rm mag,g}(\nu_{\rm max})=0.4\, \mu$Hz. \textsl{Bottom panel:} ${\rm d}\pi_0$-LS periodogram of the same rotating and magnetized star. In each panel, the star is observed with various inclination angles with respect to the rotation axis of the star following the color code in the legend. LS periodogram are zoomed around the true value of the period spacing $\Delta\Pi_1=75.2$ sec.}
    \label{fig:LSrotincl}
\end{figure}

To retrieve $\Delta\Pi_1$ starting from an artificial PSD, we have to compute a LS periodogram of the $\pi$-stretched spectrum and to search for the period corresponding to the maximum amplitude in the LS periodogram. The orange curve in the top panel of Fig.~\ref{fig:LSrotincl} is an example of the measurement of $\Delta\Pi_1$ from the LS periodogram built from Eq.~\ref{eq:deltatau_obs} for a synthetic PSD representing a non-magnetized, non-rotating star inside which mixed-mode frequencies are not affected ($\nu=\nu_{\rm obs}$). The LS method applied to this star leads to a correct measurement of $\Delta\Pi_{1_{\rm LS }}$ of $\Delta\Pi_1=75.2$ seconds. However, dynamical processes taking place inside the star can affect the frequency of the modes. In particular, wave propagation at low frequencies is affected by rotation and magnetism \citep[e.g.,][]{Ouazzani2013, Fuller2015}. In the following, we use $\tau_{{\rm obs}}$ to denote the stretched period computed from observed oscillation frequencies:%the observed mixed mode period spacing associated with ($\ell=1, m$) components as:
\begin{equation}
    \Delta\tau_{{\rm obs}_m} =\int_{\nu_{{\rm obs}_{m, n+1}}}^{\nu_{{\rm obs}_{m, n}}} {\rm d}\tau_{{\rm obs}} = \int_{\nu_{{\rm obs}_{m, n+1}}}^{\nu_{{\rm obs}_{m, n}}}-\frac{1}{\zeta}\frac{{\rm d}\nu_{{\rm obs}}}{\nu_{{\rm obs}}^2} \, .
    \label{eq:deltatau_obs}
\end{equation}

If mixed-mode frequencies are not affected by any external perturbations, then $\nu_{\rm obs}=\nu$, ${\rm d}\tau_{\rm obs}={\rm d}\pi$, and Eq.~\ref{eq:deltatau_obs} is reduced to Eq.~\ref{eq:tau}. However, ($\ell=1, m$) mixed modes might be spaced differently depending on dynamical mechanisms taking place inside the star. \textbfediting{We therefore introduced} $\Delta\tau_{{\rm obs}_m}$ \textbfediting{in Eq.~\ref{eq:deltatau_obs}}, which stands for the observed period spacing associated with modes with ($\ell=1, m$) components in a star \citep[see][]{Gehan2018}.%the observed mixed mode period spacing associated with ($\ell=1, m$) components as:
% \begin{equation}
%     \Delta\tau_{{\rm obs}_m} =\int_{\nu_{{\rm obs}_{m, n+1}}}^{\nu_{{\rm obs}_{m, n}}} -\frac{1}{\zeta}\frac{{\rm d}\nu_{{\rm obs}_m}}{\nu_{{\rm obs}_m}^2} \, .
%     \label{eq:deltatau_obs_m}
% \end{equation}

One of the dynamical mechanisms known to produce different spacings between different "m" orders is rotation. Rotation lifts the degeneracy among the frequencies of modes with the same $(n,\ell)$ but different azimuthal orders, $m$ \citep{Gizon2003, Garcia2019}. Rotation effects on mixed mode frequencies can be approximated on the first order as evolved stars with solar-like oscillations are relatively slow rotators \citep[in contrast to the case of g-modes in rapidly rotating dwarfs, see][]{Aerts2021}. We define $\delta\nu_{\rm rot}$ as the frequency perturbation induced by rotation, so that the observed frequencies are expressed as a function of the azimuthal order, $m$:%$\nu_{{(n, \ell=1, m)}_{\rm obs}}$ frequency writes:
\begin{equation}
    \nu_{{\rm obs}_m} = \nu_{m=0} + m \delta\nu_{\rm rot}\, .
\end{equation}

For mixed modes that are sensitive to the rotation rate of both the envelope and the core, we approximate $\delta\nu_{\rm rot}$ as \citep{Goupil2013a}:
\begin{equation}
    \delta\nu_{\rm rot} = \zeta \delta\nu_{\rm rot,g} +\left(1-\zeta\right)\delta\nu_{\rm rot, p}\,.    % \delta\nu_{\rm rot} = \textcolor{red}{\int_0^{\pm1}\zeta{\rm d}m} \delta\nu_{\rm rot,g} +\left(1-\zeta\right)\delta\nu_{\rm rot, p}\,.
\end{equation}
This approximation has proven to be sufficient to probe spatially-resolved rotation profiles in stellar interiors \citep[e.g.,][]{Deheuvels2014a, DiMauro2016b, Triana2017}. In addition, g-$m$ modes are mostly sensitive to the rotation rate near the core, so that the rotational splitting associated with the internal process can be approximated as:

\begin{equation}
    \delta\nu_{\rm rot} \approx \zeta \delta\nu_{\rm rot,g} \,.    % \delta\nu_{\rm rot} \approx \textcolor{red}{\int_0^{\pm1}\zeta{\rm d}m} \delta\nu_{\rm rot,g} \,.
\end{equation}

The observed g-$m$ mode frequencies in the PSD in the presence of rotation therefore depends on the azimuthal order $m$ as:
\begin{equation}
     \nu_{{\rm obs}_m} = \nu + \delta\nu_{{\rm rot}_m},
     \label{eq:nuobsm}
    % \nu_{\rm obs} = \nu + \textcolor{red}{\int_{\nu}^{\nu_{\rm obs}}\zeta{\rm d}m} \delta\nu_{\rm rot,g}
\end{equation}
\noindent with $\nu=\nu_{m=0}$ the unperturbed mixed-mode frequency and $\delta\nu_{{\rm rot}_m} = m \zeta \delta\nu_{\rm rot,g}\, $. As a result of Eq.~\ref{eq:nuobsm}, one $(n, \ell)$ mode is split into three evenly spaced components resulting from the first-order effect of rotation on oscillation frequencies. A first-order rotating synthetic spectrum with a rotation splitting $\delta\nu_{\rm rot,g}=0.69\, \mu$Hz is represented on the second panel of Fig.~\ref{fig:PSDs}.\\

By assuming that the rotational perturbation of g-$m$ modes is independent of the mixed-mode frequencies at first order (i.e. we neglect the frequency dependency of the $\zeta$ function close to g-$m$ modes, $\zeta(\nu)\approx\zeta$), we have ${\rm d}\nu_{{\rm obs}_m}\approx{\rm d}\nu$. By using Eq.~\ref{eq:tau}, Eq.~\ref{eq:deltatau_obs} can be rewritten as a function of the unperturbed ${\rm d}\pi$ as: %Eq.~\ref{eq:deltatau_obs} rewrites as:
%, and by considering

% \begin{equation}
%     \frac{{\rm d}\nu_{\rm obs} }{{\rm d}\nu}= 1 + \zeta m\delta\nu_{\rm rot,g}{{\rm d}\nu}\, ,
%     % \frac{{\rm d}\nu_{\rm obs} }{{\rm d}\nu}= 1 + \frac{\textcolor{red}{{\rm d}\int_{\nu_0}^{\nu_{\pm1}}\zeta{\rm d}m}}{{\rm d}\nu} \delta\nu_{\rm rot,g}
% \end{equation}
\begin{equation}
    \Delta\tau_{{\rm obs}_m} = \int \frac{{\rm d}\pi}{\left(1+\frac{\displaystyle \delta\nu_{{\rm rot}_m}}{\displaystyle\nu}\right)^2}\, .
    % {\rm d}\tau_{{\rm obs}_m} = \frac{{\rm d}\tau}{\left(1+m\textcolor{red}{\int_0^{\pm1}\zeta{\rm d}m}\frac{\displaystyle \delta\nu_{\rm rot}}{\displaystyle\nu}\right)^2}\, .
    \label{eq:deltatau_rot}
\end{equation}
% \begin{equation}
%     \Delta\tau_{{\rm obs}_m} = \int \frac{{\rm d}\pi}{\left(1+m\zeta\frac{\displaystyle \delta\nu_{\rm rot, g}}{\displaystyle\nu}\right)^2}\, .
%     % {\rm d}\tau_{{\rm obs}_m} = \frac{{\rm d}\tau}{\left(1+m\textcolor{red}{\int_0^{\pm1}\zeta{\rm d}m}\frac{\displaystyle \delta\nu_{\rm rot}}{\displaystyle\nu}\right)^2}\, .
%     \label{eq:deltatau_rot}
% \end{equation}
As a result, the period spacing of pure g-modes defined from Eq.~\ref{eq:dpi} can be measured from the distance between $m=0$ modes with consecutive radial order $n$ expressed in the stretch variable $\tau_{{\rm obs}}$ even in the rotating case. The LS method described in Section~\ref{sec:measure_dpi} can thus still be computed from the stretched PSD (Eq.~\ref{eq:deltatau_obs}) in the rotating case to retrieve $\Delta\Pi_1$, because $m=0$ modes will maintain the signal of unperturbed g-$m$ mixed-modes in the LS periodogram.%and $\Delta\tau_{\rm obs}\not= \Delta\Pi_1$.

Previous studies \citep{Vrard2016, Gehan2018} therefore neglected the effect of rotation on the measurement of $\Delta\Pi_1$. Indeed, as long as the $m=0$ component (that is not affected by rotation) is visible in the PSD, the methodology (described in Sect.~\ref{sec:measure_dpi}) is still valid and the effect of rotation on the measured $\Delta\Pi_1$ can be neglected \citep[we refer the reader to][for more details about the effect of rotation on $\Delta\Pi_1$ measurement when the star is not observed pole-on]{Vrard2016}. 
However, if the star is observed edge-on (inclination angle close to $90 \degree$), the non-perturbed frequency of mixed modes (associated with $m=0$ modes) is no longer present in the PSD. As a result, for stars observed with high inclination ($i\geq 45\degree$) the stretching of the PSD according to Eq.~\ref{eq:deltatau_obs} may not reveal the periodicity of mixed modes from the data. \\ %and can therefore not be precisely measured by this method. 

 Such a phenomenon is represented on the top panel of Fig.~\ref{fig:LSrotincl}. It shows the LS periodogram computed for a typical RGB star, with a rotation signature, ($\delta\nu_{\rm rot}=0.69\, \mu$Hz), and observed with various stellar inclinations ($i$). We show that for $i\lessapprox 45\degree$, the measurement of $\Delta\Pi_1$ is not affected by rotation as only the amplitude (not the period) of the main detected peak is altered. However, for larger inclination angles, the impact of the $m=-1,1$ components dominate the LS periodogram. As demonstrated by the dependency in $\left(1+\frac{\delta\nu_{{\rm rot}_m}}{\nu}\right)^{-2}$ in Eq.~\ref{eq:deltatau_rot}, the $m=1$ (and $=-1$) periods are not evenly spaced in the commonly used ${\rm d}\tau_{{\rm obs}_{0}}$-stretched spectrum, leading to a spreading of the power density around non-resolved $m=1$ and $m=-1$ peaks in the LS periodogram around $\Delta\Pi_1$.
 
As presented in Section~\ref{sec:measure_dpi}, the $\delta\left(\Delta\Pi_{1_{\rm over}}\right)$ precision obtained on $\Delta\Pi_1$ strongly depends on the amplitude of the peak in the LS periodogram. When the amplitude goes to zero in the LS periodogram as the inclination increases towards $|i|\approx 90\degree$, the uncertainty on the measurement of $\Delta\Pi_1$ goes to infinity. We conclude that the measurement of $\Delta\Pi_1$ with this LS method is robust only for rotating stars with relatively low inclination angles ($i \leq 45\degree$). %We focus in the rest of the article on stars with $i \leq 45\degree$ To ensure that $\Delta\Pi_1$ is properly estimated in rotating stars from the period associated with the maximum of the LS.

\section{Effect of internal magnetism on the measurement of g-mode period spacing}
\label{sec:mageffecttheory}
\subsection{Magnetic field topology and amplitude inside evolved stars with solar-like oscillations}

\label{sec:field_topo}

Evolved stars with solar-like oscillations that are considered in this work are descendants of low- and intermediate-mass main sequence stars, which have themselves evolved from fully convective bodies during the PMS. Convective zones are known to generate and to host stochastic dynamo fields \citep{Brun2017a}. Over the past 20 years, many studies have investigated how such dynamo-generated fields might remain trapped into the radiative interior once the convection has come to an end inside the star during the MS \citep[e.g.,][]{Braithwaite2004, Braithwaite2006a, Braithwaite2008a, Arlt2014, Emeriau-Viard2017}. \cite{Braithwaite2004} were the first to simulate the relaxation of completely stochastic fields in a non-rotating radiative interior. They show that dynamo-generated fields can become stabilized when the convection process ends. From these simulations, we deduce that initial stochastic dynamo fields placed into a stably stratified interior might relax into stable, large-scale, mixed poloidal and toroidal magnetic field configurations \citep{Braithwaite2008a}. This inverse turbulent cascade mechanism is described theoretically in \cite{Duez2010}. Purely toroidal and purely poloidal magnetic configurations are known to be unstable \citep[e.g.,][]{Tayler1973, Markey1973, Braithwaite2006, Braithwaite2007a}. Combining toroidal and poloidal components ensures the stability of the field \citep{Tayler1980, Braithwaite2009}. Once a stable field is formed in the radiative interior, the characteristic Ohmic timescale on which the field gets dissipated is of about $10^{10}$ years, much longer than the lifetime of the star. Hence, if stable magnetic fields are formed inside the radiative interior during the main sequence, they would remain trapped inside the radiative interior on the RGB if no external dissipating mechanisms are at work.\\

\begin{figure}
    \centering
    \includegraphics[width=1\linewidth]{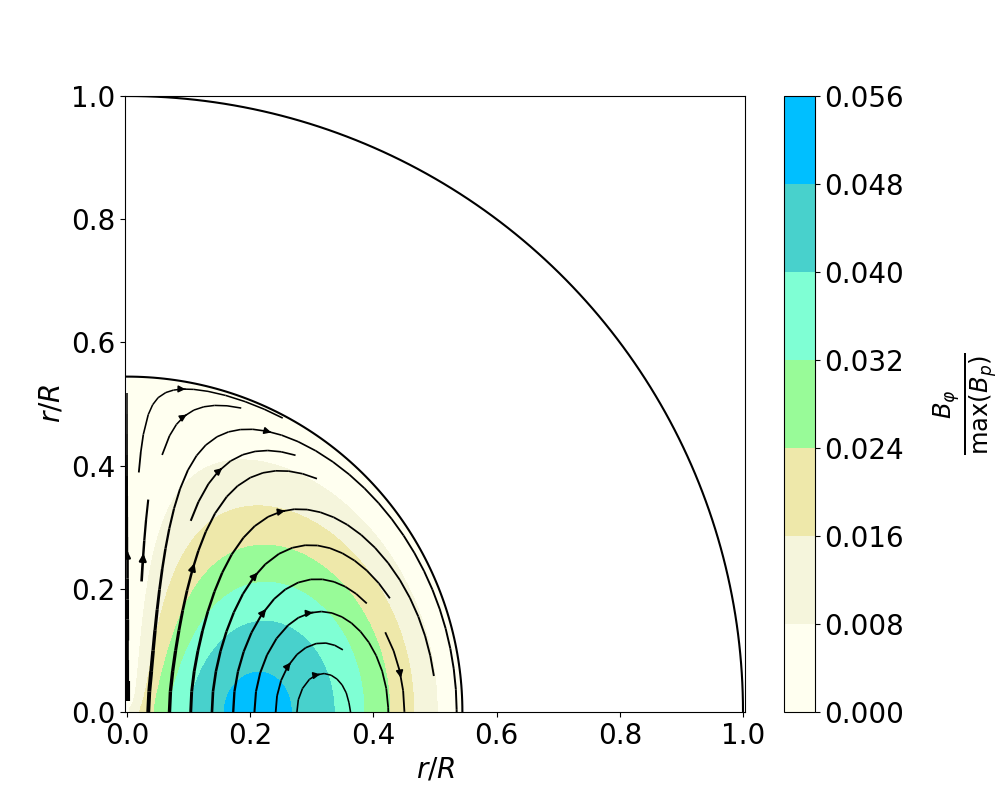}%from C02DL05QMD6V:magperturb lbugnet$ python3 -i compute_shifts_autosave_with_metallicity_with_differential_rotation.py 1.5 0.02 450 451 1
    \caption{Magnetic field configuration inside a $M=1.5\,{\rm M}_\odot$, $Z=0.02$, $2.71$ Gyr evolved star with solar-like oscillations on the RGB. The field is confined within the inner black line (convective-radiative boundary) and the stellar surface is indicated by the outer black line.}
    \label{fig:Bfieldtopo}
\end{figure}

\textbf{Field topology:}
The topology and energy of stable mixed poloidal and toroidal field has been investigated numerically by \cite{Braithwaite2008a}, and semi-analytically by \cite{Akgun2013}. \cite{Duez2010} provided a semi-analytic description of the stable mixed toroidal and poloidal fields similar to those resulting from the simulations of \cite{Braithwaite2008a}:
\begin{equation}
    \boldsymbol{B}\left(r, \theta\right)=
    \displaystyle
    \frac{1}{r \sin{\theta}}\left( \boldsymbol{\nabla} \psi(r,\theta) \wedge \boldsymbol{e_\varphi} + \lambda \frac{\psi(r,\theta)}{R} \boldsymbol{e_\varphi}\right) \, ,
    \label{eq:compu}
\end{equation}
\noindent where $\psi$ is the stream function satisfying 
\begin{equation}
    \psi(r,\theta)=\mu_0 \alpha \lambda \frac{A(r)}{R}\sin^2{\theta}\, ,
\end{equation} with $\mu_0$ as the vacuum magnetic permeability, $\alpha$ as a normalization constant fixed by the chosen magnetic-field amplitude, $\lambda$ as the eigenvalue of the problem to be determined, $R$ as the radius of the star, and 
\begin{multline}
A(r)=-r j_1\left(\lambda \frac{r}{R} \right) \int_r^{R} y_1\left(\lambda \frac{x}{R} \right)\rho x^3 \textrm{d}x\\
-r y_1\left(\lambda \frac{r}{R} \right) \int_0^r j_1\left(\lambda \frac{x}{R} \right)\rho x^3 \textrm{d}x,
\label{eq:A}
\end{multline}
\noindent with $j_1$ ($y_1$) as the first-order spherical Bessel function of the first (second) kind \citep{Abramowitz1972} and $\rho$=$\rho(r)$ the hydrostatic density profile of the star. To confine the field inside the radiative interior, we set $\lambda$ as the smallest positive value for $\boldsymbol{B}$ to vanish at the convective-radiative boundary. The resulting magnetic field configuration for a typical RG ($M=1.5\,{\rm M}_\odot$, $Z=0.02$, $2.71$ Gyr) is represented in Fig.~\ref{fig:Bfieldtopo}.\\

\textbf{Field amplitude:}
We cannot (thus far) detect buried magnetic fields through direct observations and so, we can only try to predict their amplitude. Considering the fossil field scenario described above, and assuming that the magnetic flux is conserved inside the mass coordinate of the radiative interior as the star evolves from the MS, \cite{Bugnet2021} showed that the magnetic field amplitude inside red giants should be in the range $[10^5: 10^7]$ Gauss \citep[depending on the efficiency of the dynamo action, see][]{Augustson2019}. We refer to \cite{Bugnet2021} for details about the conservation of the magnetic flux along the evolutionary process resulting in these numbers.

\subsection{Magnetic effect on mixed-mode frequencies}
The effect of stable fossil internal magnetic fields on the frequencies of solar-like mixed oscillations has been investigated by \cite{Bugnet2021}, \cite{Mathis2021}, and \cite{Loi2021}. By performing a first-order perturbative analysis on the eigenfrequencies of the modes, \cite{Bugnet2021} showed that axisymmetric fossil magnetic fields lifts the degeneracy on the azimuthal order of mixed modes and generates an asymmetric shift in the observed frequencies. The authors derived the complete first-order perturbation and demonstrated, through the use of the pulsation code GYRE \citep{Townsend2013}, that such magnetic signature on the mixed-mode frequency pattern might be detectable in the \textsl{Kepler} data. \cite{Mathis2021} derived asymptotic expressions for the magnetic signature on mixed-mode frequencies corresponding to the dominant effect revealed by \cite{Bugnet2021}. Finally, \cite{Mathis2021} study showed that the effect of a mixed poloidal and toroidal field aligned with the rotation axis of the star (as represented in Fig.~\ref{fig:Bfieldtopo}) on g-$m$ mixed-mode frequencies can be approximated by
\begin{align}
{\delta\nu}_{{\rm mag,g}_m}(\nu)=\frac{1}{2\left(2\pi\right)^4}\frac{{\rm B}_0^2}{4\pi{\rho}_{c}R_{\rm rad}^2\nu}\frac{N_{\rm max}^2}{\nu^2}&\ell\left(\ell+1\right){\mathcal C}_{\ell,m}\left(\nu\right)\nonumber\\
&\times \frac{\displaystyle{\int_0^1}\displaystyle{\frac{b_r^2{\widehat N}^2}{\left(\rho/\rho_c\right)x^2}}{\widehat N}\displaystyle{\frac{{\rm d}x}{x}}}{\displaystyle{\int_0^1}{\widehat N}\displaystyle{\frac{{\rm d}x}{x}}},
\label{eq:Mathis}
\end{align}

% \begin{equation}
% {\delta\nu}_{\rm g}=A\frac{{\rm B}_0^2}{\nu^3},
% \label{eq:Mathis}
% \end{equation}
% with
%  \begin{equation}
% A=\frac{1}{2}\frac{1}{4\pi{\rho}_{c}R_{\rm rad}^2}{N_{\rm max}^2} \Lambda_{k,m}\left(\nu\right){\mathcal C}_{k,m}\left(\nu\right)\frac{\displaystyle{\int_0^1}\displaystyle{\frac{b_r^2{\widehat N}^2}{\left(\rho/\rho_c\right)x^2}}{\widehat N}\displaystyle{\frac{{\rm d}x}{x}}}{\displaystyle{\int_0^1}{\widehat N}\displaystyle{\frac{{\rm d}x}{x}}},
% \label{eq:Mathis}
% \end{equation}

\noindent where we introduce:
\begin{inparadesc}
    \item[1.] the dimensionless radius $x=r/R_{\rm rad}$ inside the radiative interior, with $R_{\rm rad}$ the radius of the radiative zone,
    \item[2.] the ratio of the dimensionless density profile $\rho$ inside the radiative interior and of the central density, $\rho_c$,
    \item[3.]  the maximum of the Brunt-V\"ais\"al\"a frequency, $N_{\rm max}$, and its dimensionless radial profile, ${\widehat N}=N(r)/N_{\rm max}$, and
    \item[4.] the magnetic field amplitude along the field lines, ${\rm B}_0$, and its dimensionless radial profile, $b_r$
    % \item \Lambda_{k,m}\left(\nu\right)$ the horizontal eigenvalues of the Hough functions \citep[e.g.][]{Hough1898,Lee1997,Townsend2003} $\left\{H_r\left(\cos\theta\right), \partial_\theta Y_\ell^m e^{-im\varphi}\left(\cos\theta\right),\left( \frac{m}{\sin\theta} Y_\ell^m\right)\left(\cos\theta\right)\right\}$ that generalize the spherical harmonics when taking into account the rotation within the TAR (we refer the reader to the Appendix A of \cite{Mathis2021} for the details of their definition),
    %   \item 
    \item and the angular integral:
%     \begin{equation}
% {\mathcal C}_{k,m}\left(\nu\right)=\frac{\displaystyle{\int_{0}^{\pi}\left[\partial_\theta Y_\ell^m e^{-im\varphi}^{2}\left(\cos\theta\right)+\left( \frac{m}{\sin\theta} Y_\ell^m\right)^{2}\left(\cos\theta\right)\right]\cos^2\theta\sin\theta{\rm d}\theta}}{\displaystyle{\int_{0}^{\pi}\left[\partial_\theta Y_\ell^m e^{-im\varphi}^{2}\left(\cos\theta\right)+\left( \frac{m}{\sin\theta} Y_\ell^m\right)^{2}\left(\cos\theta\right)\right]\sin\theta{\rm d}\theta}}\, .
% \end{equation}
%     \begin{equation}
% {\mathcal C}_{\ell,m}\left(\nu\right)=\frac{\displaystyle{\int_{0}^{\pi}\left[\left(\partial_\theta Y_\ell^m \right)^{2}+\left( \frac{m}{\sin\theta} Y_\ell^m\right)^{2}\right]\cos^2\theta\sin\theta{\rm d}\theta}}{\displaystyle{\int_{0}^{\pi}\left[\left(\partial_\theta Y_\ell^m\right)^{2}+\left( \frac{m}{\sin\theta} Y_\ell^m\right)^{2}\right]\sin\theta{\rm d}\theta}}\, .
% \end{equation}
\begin{equation}
C_{l,m}=\frac{\displaystyle{\int_{0}^{\pi}}\displaystyle{\left[\left|\cos\theta\partial_{\theta}Y_l^m\right|^2+m^2\left|\frac{\cos\theta}{\sin\theta}Y_l^m\right|^2\right]\sin\theta{\rm d}\theta}}{l\left(l+1\right)}.
\end{equation}
\end{inparadesc}

This prescription of the magnetic signature on dipolar mixed-mode frequencies is key for a search for internal magnetic fields inside the radiative interior of evolved stars.
%As demonstrated by \cite{Bugnet2021} and \cite{Mathis2021}, the perturbation induced by a stable axisymmetric magnetic field inside the radiative interior on dipolar mixed-mode frequencies can be expressed as 

From Eq.~\ref{eq:Mathis}, we deduce the expression of the internal magnetic perturbation on mixed-mode frequencies for each of the $m$ azimuthal degree as:
\begin{equation}
    \delta\nu_{{\rm mag}_m}(\nu)= m^*\zeta \delta\nu_{{\rm mag, g}}(\nu)=m^*\zeta\frac{A{\rm B}_0^2 }{\nu^3}\, ,
    \label{eq:dnumag}
\end{equation}
with $m^*=\frac{\displaystyle|m|+1}{\displaystyle2}$, $A$ as the normalization factor depending on the magnetic field topology and of the structure of the star containing all remaining terms in Eq.~\ref{eq:Mathis} \citep[see][for more details]{Mathis2021}, ${\rm B}_0$ as the magnetic field amplitude, and $\delta\nu_{{\rm mag, g}}$ as the magnetic effect associated with pure ${\rm g}$-modes. As a result, the magnetic perturbation is a function of the frequency, and the perturbation on the $m=1$ and $m=-1$ mixed-mode components are equal to twice the perturbation on the $m=0$ component. The magnetic signature is therefore very different from the rotational signature, as all the $m$ components are affected and form a non-symmetric pattern in frequency \citep[see][]{Bugnet2021}.
% \sout{the magnetic perturbation strongly depends on the unperturbed frequency of the mode}

% \textbf{Effect of the mass and metallicity on the perturbation by magnetic fields}

% \textbf{General law for magnetic effects on mixed-mode frequencies}

% \begin{equation}
%     \left(\delta\nu\right)_{\rm g} \propto \frac{A{\rm B}_0^2}{\nu^3}
% \end{equation}

\subsection{Error of the measurement of $\Delta\Pi_1$ in the presence of magnetic fields}\label{sec:mag}
 We now consider a non-rotating star and an internal axisymmetric magnetic field to investigate its effect on the LS methodology to estimate $\Delta\Pi_1$. In the presence of magnetism, the observed $(n, \ell=1, m)$ frequencies write as the following function of the unperturbed mixed-mode frequency:
% \subsubsection{Skewness of the peak in LS}

% \subsection{$\Delta\Pi_1$ estimates in presence of rotation}

% \subsection{$\Delta\Pi_1$ estimates in presence of magnetism}

% \section{Introducing $\pi$, the stretched period in presence of magnetism}
% The observed period spacings of mixed modes can again be written as:
% \begin{equation}
%     {\rm d}\tau_{\rm obs} = -\frac{1}{\zeta}\frac{{\rm d}\nu_{\rm obs}}{\nu_{\rm obs}^2} \, .
% \end{equation}

% \noindent However, in presence of magnetism;
\begin{equation}
    \nu_{{\rm obs}_m}=\nu + \delta\nu_{{\rm mag}_m}(\nu)
    \label{eq:nuobsmag}
,\end{equation}
% \begin{equation}
%     \nu_{{\rm obs}_m}=\nu + m^*\zeta \frac{A {\rm B}_0^2}{\nu^3}
%     \label{eq:nuobsmag}
% \end{equation}
so that
\begin{equation}
    \frac{{\rm d}\nu_{{\rm obs}_m}}{{\rm d}\nu} = 1 - 3\frac{\delta\nu_{{\rm mag}_m}(\nu)}{\nu}%\textcolor{red}{+m^*\frac{A{\rm B}_0^2}{\nu^3}\frac{{\rm d}\zeta}{{\rm d}\nu}}\, .
    \label{eq:dnu}
.\end{equation}
% \begin{equation}
%     \frac{{\rm d}\nu_{{\rm obs}_m}}{{\rm d}\nu} = 1 - 3m^*\zeta \frac{A{\rm B}_0^2}{\nu^4}%\textcolor{red}{+m^*\frac{A{\rm B}_0^2}{\nu^3}\frac{{\rm d}\zeta}{{\rm d}\nu}}\, .
%     \label{eq:dnu}
% \end{equation}

% In the following study, we write ${\rm d}\pi_{{\rm obs}_m}$ the observed $\pi$-stretched period associated with modes with ($\ell=1, m$) components in a magnetized star as:%the observed mixed mode period spacing associated with ($\ell=1, m$) components as:
% \begin{equation}
%     {\rm d}\pi_{{\rm obs}_m} = -\frac{1}{\zeta}\frac{{\rm d}\nu_{{\rm obs}_m}}{\nu_{{\rm obs}_m}^2} \, .
%     \label{eq:deltatau_obs}
% \end{equation}

\noindent In the case of magnetism only, we obtain an equation analog to Eq.~\ref{eq:deltatau_rot}:
% \noindent with $m^*=\frac{\displaystyle|m|+1}{\displaystyle2}$,
% \begin{equation}
%     {\rm d}\tau_{{\rm obs}_m} = \frac{1}{\zeta}\frac{\displaystyle1 - 3m^*\zeta \frac{A}{\nu^4} {\rm d}\nu}{\displaystyle\left(\nu+m^*\zeta \frac{\displaystyle A}{\nu^3}\right)^2} \, .
% \end{equation}

\begin{equation}
\Delta\tau_{{\rm obs}_m} \approx \Delta\Pi_1 \frac{1 - 3 \displaystyle\frac{\displaystyle \delta\nu_{{\rm mag}_m}(\nu)}{\displaystyle \nu} }{\displaystyle\left(1+\frac{\delta\nu_{{\rm mag}_m}(\nu)}{\nu}\right)^2} \,.%-\frac{{\rm d}\nu}{\displaystyle\zeta \nu^2}\frac{1 - 3m^*\zeta \frac{\displaystyle A{\rm B}_0^2}{\displaystyle \nu^4} }{\displaystyle\left(1+m^*\zeta \frac{A{\rm B}_0^2}{\nu^4}\right)^2} \,.
    % {\rm d}\tau_{{\rm obs}_m} = -\frac{{\rm d}\nu}{\displaystyle\zeta \nu^2}\frac{1 - 3m^*\zeta \frac{\displaystyle A{\rm B}_0^2}{\displaystyle \nu^4} \textcolor{red}{+m^*\frac{A{\rm B}_0^2}{\nu^3}\frac{{\rm d}\zeta}{{\rm d}\nu}}}{\displaystyle\left(1+m^*\zeta \frac{A{\rm B}_0^2}{\nu^4}\right)^2} \, .
    \label{eq:taummag}
\end{equation}

% \begin{equation}
% \Delta\tau_{{\rm obs}_m} \approx \Delta\Pi_1 \frac{1 - 3m^*\zeta \frac{\displaystyle A{\rm B}_0^2}{\displaystyle \nu^4} }{\displaystyle\left(1+m^*\zeta \frac{A{\rm B}_0^2}{\nu^4}\right)^2} \,.%-\frac{{\rm d}\nu}{\displaystyle\zeta \nu^2}\frac{1 - 3m^*\zeta \frac{\displaystyle A{\rm B}_0^2}{\displaystyle \nu^4} }{\displaystyle\left(1+m^*\zeta \frac{A{\rm B}_0^2}{\nu^4}\right)^2} \,.
%     % {\rm d}\tau_{{\rm obs}_m} = -\frac{{\rm d}\nu}{\displaystyle\zeta \nu^2}\frac{1 - 3m^*\zeta \frac{\displaystyle A{\rm B}_0^2}{\displaystyle \nu^4} \textcolor{red}{+m^*\frac{A{\rm B}_0^2}{\nu^3}\frac{{\rm d}\zeta}{{\rm d}\nu}}}{\displaystyle\left(1+m^*\zeta \frac{A{\rm B}_0^2}{\nu^4}\right)^2} \, .
%     \label{eq:taummag}
% \end{equation}

% The magnetized $m=0$ frequency pattern stretched through the use of Eq.~\ref{eq:tau} presents spacings following : \begin{equation}
%     {\rm d}\tau_{{\rm obs}_0} = -\frac{{\rm d}\nu}{\displaystyle\zeta \nu^2}\frac{1 - \frac{\displaystyle 3\zeta}{\displaystyle 2} \frac{\displaystyle A {\rm B}_0^2}{\displaystyle \nu^4} }
%     {\displaystyle\left(1+\frac{\zeta}{2} \frac{A {\rm B}_0^2}{\nu^4}\right)^2} \, .\label{eq:tau0mag}
    % {\rm d}\tau_{{\rm obs}_0} = -\frac{{\rm d}\nu}{\displaystyle\zeta \nu^2}\frac{1 - \frac{\displaystyle 3\zeta}{\displaystyle 2} \frac{\displaystyle A {\rm B}_0^2}{\displaystyle \nu^4} \textcolor{red}{+m^*\frac{A{\rm B}_0^2}{\nu^3}\frac{{\rm d}\zeta}{{\rm d}\nu}}}
    % {\displaystyle\left(1+\frac{\zeta}{2} \frac{A {\rm B}_0^2}{\nu^4}\right)^2} \, .\label{eq:tau0mag}
% \end{equation}
As $\Delta\tau_{{\rm obs}_m}\neq \Delta\Pi_1$ %${\rm d}\tau_{{\rm obs}_0}\neq {\rm d}\pi$ 
for all $m \in [-1, 0, 1]$ values through Eq.~\ref{eq:taummag}, the classical stretching made by using Eq.~\ref{eq:deltatau_obs} should not result in an accurate measurement of $\Delta\Pi_1$ for any of the $m$ orders, as opposed to the rotational case where the $m=0$ modes remain evenly spaced. Equation~\ref{eq:taummag} therefore demonstrates that the use of the ${\rm d}\tau_{{\rm obs}}$ period stretching (see Eq.~\ref{eq:deltatau_obs}), as developed by \cite{Vrard2016}, leads to biased estimates of $\Delta\Pi_1$ if the radiative interior of the star is magnetized, because ${\rm d}\tau_{{\rm obs}} ={\rm d}\tau_{{\rm obs}_0}\neq {\rm d}\pi$. 

In the third panel of Fig.~\ref{fig:PSDs} we represent the PSD of a magnetized star with $\delta\nu_{\rm mag}(\nu_{\rm max})=50\delta f_{\rm res}$, considering an inclination angle of $i=45\degree$.
The ${\rm d}\tau_{\rm obs}$-LS periodogram of such a PSD is represented in the second panel of Fig.~\ref{fig:LSrotincl} following Eq.~\ref{eq:deltatau_obs}, along with other LS periodogram computed from artificial data corresponding to various observation angles. As expected, stretching the PSD following Eq.~\ref{eq:deltatau_obs} does not allow for an accurate measurement of $\Delta\Pi_1$ when the star is magnetized. The erroneous estimation of $\Delta\Pi_1$ increases with the inclination angle, as $m=1$ and $m=-1$ modes, which are the most affected by magnetic fields, tend to dominate the spectrum when $i$ increases.
To correct the measurement of $\Delta\Pi_1$, we adapt the stretch function to account for the true period of g-mode $\pi,$ instead of $\tau_{{\rm obs}}$:% so that the estimate of $\Delta\Pi_1$ is accurate when the star is magnetized:
% \begin{equation}
%     % \Delta\Pi_1 = \int_{{\nu_{\rm obs}_m}} {\rm d}\pi_m\, .
%         \Delta\Pi_1 = \int_{{\nu_n}}^{\nu_{n+1}} -\frac{{\rm d}\nu}{\displaystyle\zeta \nu^2} = \int_{{\nu_n}}^{\nu_{n+1}} {\rm d}\pi\, .
% \end{equation}

\noindent From Eq.~\ref{eq:taummag} we deduce that the $\pi$ function can be written as a function of one of the $m=1$, $m=0$, or $m=1$ mode frequencies, following:
% \begin{equation}
%     {\rm d}\tau_{{\rm obs}_m} = {\rm d}\tau\frac{\displaystyle1 - 3m^*\zeta \frac{A {\rm B}_0^2}{\nu^4} }{\displaystyle\left(1+m^*\zeta \frac{A {\rm B}_0^2}{\nu^4}\right)^2} \, .
% \end{equation}
\begin{equation}
    {\rm d}\pi = -\frac{1}{\zeta}\frac{{\rm d}\nu_{{\rm obs}_m}}{\nu_{{\rm obs}_m}^2} \frac{\displaystyle\left(1+ \frac{\delta\nu_{{\rm mag}_m}(\nu)}{\nu}\right)^2}{\displaystyle1 - 3 \frac{\delta\nu_{{\rm mag}_m}(\nu)}{\nu} } \, .
    \label{eq:dpi_mag_nuobs}
\end{equation}
% \begin{equation}
%     {\rm d}\pi = -\frac{1}{\zeta}\frac{{\rm d}\nu_{{\rm obs}_m}}{\nu_{{\rm obs}_m}^2} \frac{\displaystyle\left(1+m^*\zeta \frac{A {\rm B}_0^2}{\nu^4}\right)^2}{\displaystyle1 - 3m^*\zeta \frac{A {\rm B}_0^2}{\nu^4} } \, ,
%     \label{eq:dpi_mag_nuobs}
% \end{equation}

% \subsection{Corrected measure of $\Delta\Pi_1$ in presence of magnetism}

\subsection{Combined effects of rotation and magnetism on the measurement of $\Delta\Pi_1$}

The artificial PSD of a RG in the presence of rotation ($\delta\nu_{\rm rot, g}=0.69\, \mu$Hz) and an internal stable magnetic field ($\delta\nu_{\rm mag, g}(\nu_{\rm max})=0.4\, \mu$Hz) observed from an angle $i=45\degree$ is represented in the bottom panel of Fig.~\ref{fig:PSDs} for reference. As the star is magnetized, the use of Eq.~\ref{eq:deltatau_obs} does not lead to an accurate value of $\Delta\Pi_1$ with the ${\rm d}\tau_{\rm obs}$-LS method, as represented in the third panel of Fig.~\ref{fig:LSrotincl} for various inclination angles. In the presence of both magnetism and rotation, observed frequencies rewrite at first order as:
% \begin{equation}
%     \nu_{{\rm obs}_m}=\nu+\delta\nu_{{\rm mag}_m}+m\zeta\delta\nu_{\rm rot}\, ,
%     \label{eq:nuobsrotmag}
% \end{equation}

\begin{equation}
    \nu_{{\rm obs}_m}=\nu+\delta\nu_{{\rm mag}_m}(\nu)+\delta\nu_{{\rm rot}_m}\, ,
    \label{eq:nuobsrotmag}
\end{equation}

% \begin{equation}
%     \nu_{{\rm obs}_m}=\nu+m^*\zeta \frac{A{\rm B}_0^2}{\nu^3}+m\zeta\delta\nu_{\rm rot}\, ,
%     \label{eq:nuobsrotmag}
% \end{equation}
\noindent and $\displaystyle \frac{{\rm d}\nu_{{\rm obs}_m}}{{\rm d}\nu}$ still follows Eq.~\ref{eq:dnu}. Therefore, the relation between the period $\pi$ adequate to measure the period spacing, $\Delta\Pi_1$, and the observed period of modes with the same $m$ azimuthal order $\tau_{{\rm obs}_m}$ is:

\begin{equation}
    {\rm d}\pi = -\frac{1}{\zeta}\frac{{\rm d}\nu_{{\rm obs}_m}}{\nu_{{\rm obs}_m}^2} \frac{\displaystyle\left(1+ \frac{\delta\nu_{{\rm mag}_m}(\nu)}{\nu}+\frac{\delta\nu_{{\rm rot}_m}}{\nu}\right)^2}{\displaystyle1 - 3 \frac{\delta\nu_{{\rm mag}_m}(\nu)}{\nu} } \, .
    \label{eq:dpi_mag_and_rot_nuobs}
\end{equation}
% \begin{equation}
%     {\rm d}\tau_{m=1} = -\frac{1}{\zeta}\frac{{\rm d}\nu_{\rm obs}}{\nu_1} \frac{1}{\displaystyle1 - 3\zeta \frac{A}{\nu_1^4} } \, .
% \end{equation}

% \begin{equation}
%     {\rm d}\tau_{m=-1} = -\frac{1}{\zeta}\frac{{\rm d}\nu_{\rm obs}}{\nu_{-1}} \frac{1}{\displaystyle1 - 3\zeta \frac{A}{\nu_{-1}^4} } \, .
% \end{equation}

% \begin{equation}
%     {\rm d}\tau_{m=0} = -\frac{1}{\zeta}\frac{{\rm d}\nu_{\rm obs}}{\nu_0} \frac{1}{\displaystyle1 - \frac{3\zeta}{2} \frac{A}{\nu_0^4} } \, .
% \end{equation}

\section{Ways to probe the presence of internal magnetism from the measurement of $\Delta\Pi_1$}
\label{sec:probeB}
% \section{Detection of magnetic effects on mixed-mode frequencies}
\subsection{The ${\rm d}\pi_0-$LS method for measuring $\Delta\Pi_1$ from axisymmetric mixed modes}
% \subsection{The importance of $m=0$ mixed modes for an accurate measure of $\Delta\Pi_1$ in the presence of magnetism: the ${\rm d}\pi_0$-LS method}
% : effect of rotation on $\pi$}

Equations~\ref{eq:dpi_mag_nuobs} and \ref{eq:dpi_mag_and_rot_nuobs} can both be rewritten, independently of the presence of rotation, as:
% \begin{equation}
%     {\rm d}\pi = -\frac{1}{\zeta}\frac{{\rm d}\nu_{{\rm obs}_m}}{\nu^2} \frac{1}{\displaystyle1 - 3m^*\zeta \frac{A{\rm B}_0^2}{\nu^4} } \, .
%     \label{eq:dpi_mag_nu}
% \end{equation}

\begin{equation}
    {\rm d}\pi = -\frac{1}{\zeta}\frac{{\rm d}\nu_{{\rm obs}_m}}{\nu^2} \frac{1}{\displaystyle1 - 3 \frac{\delta\nu_{{\rm mag}_m}(\nu)}{\nu} } \, .
    \label{eq:dpi_mag_nu}
\end{equation}

The measurement of $\Delta\Pi_1$ therefore depends on the magnetic field amplitude, on the inner structure of the stars through the dependency on $A$, and on the unperturbed frequency of mixed modes. To retrieve $\Delta\Pi_1$ from the integration of Eq.~\ref{eq:dpi_mag_nu}, only modes with a shared $m$ order must be considered simultaneously. As the $m=1$ and $m=-1$ components cannot be independently observed for stochastically-excited solar-like oscillations \citep{Gizon2003}, it is convenient to focus on $m=0$ modes. 
% \begin{equation}
%     \Delta\Pi_1 = -\int_{n, \ell=1,m=0}^{{n+1, \ell=1,m=0}} {\rm d}\pi =-\int_{n, \ell=1,m=0}^{{n+1, \ell=1,m=0}} \frac{1}{\zeta}\frac{{\rm d}\nu_{{\rm obs}}}{\nu^2} \frac{1}{\displaystyle1 - \frac{3\zeta}{2} \frac{A {\rm B}_0^2}{\nu^4} } \, .\label{eq:oes notpi_mag}
%     % \Delta\Pi_1 = -\int_{\nu_{{\rm obs}_0}} {\rm d}\pi_0 =-\int_{\nu_{{\rm obs}_0}} \frac{1}{\zeta}\frac{{\rm d}\nu_{{\rm obs}}}{\nu^2} \frac{1}{\displaystyle1 - \frac{3\zeta}{2} \frac{A {\rm B}_0^2}{\nu^4} } \, .
% \end{equation}
\begin{figure}[t]
    \centering
    \includegraphics[width=0.5\textwidth]{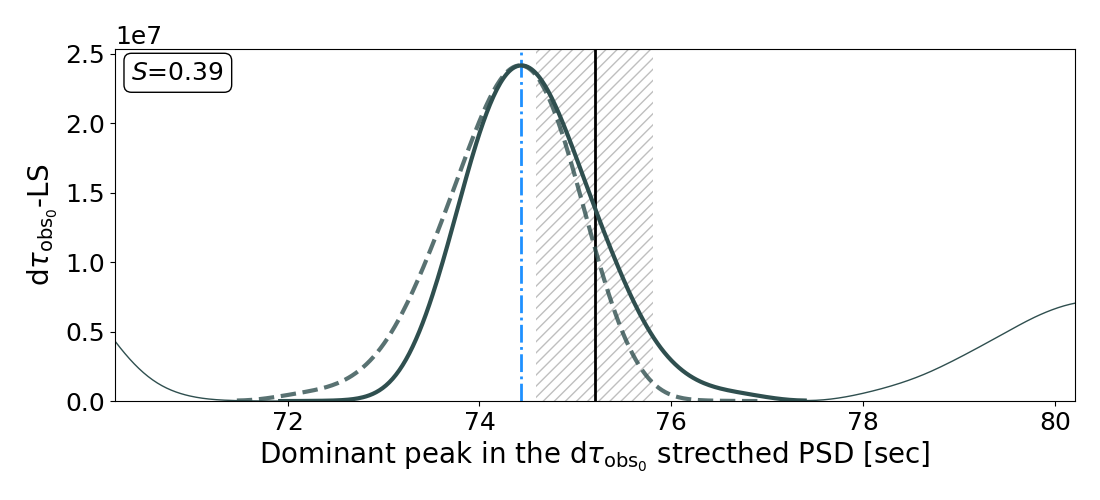}
    \includegraphics[width=0.5\textwidth]{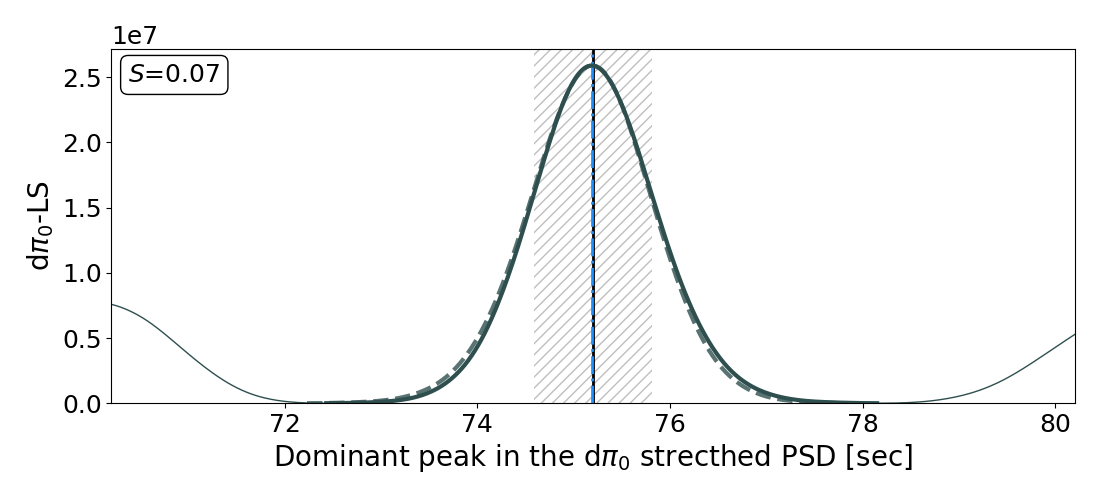}
    \includegraphics[width=0.5\textwidth]{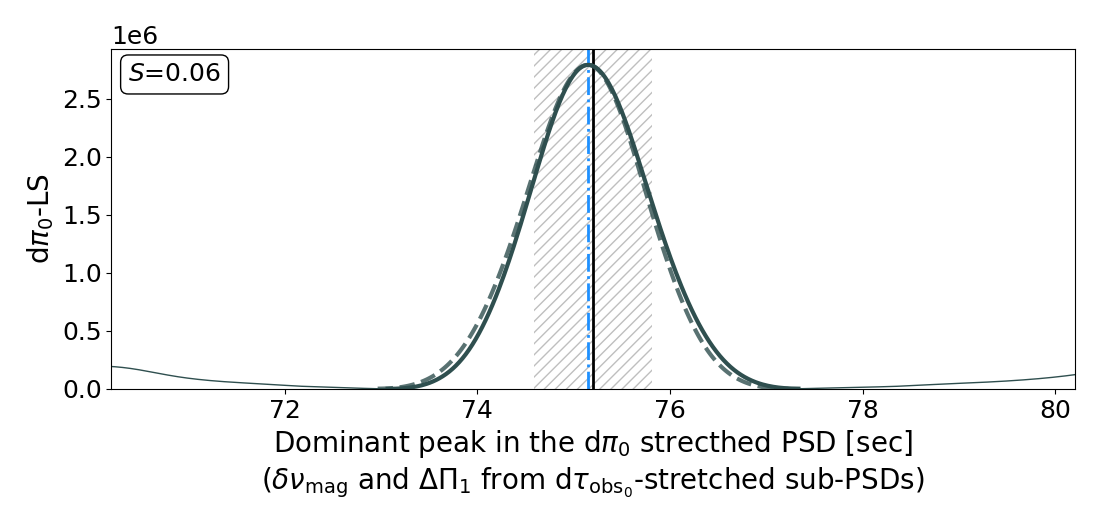}
    \caption{\textbfediting{LS periodograms for simulated stars.} \textsl{Top panel:} LS periodogram of the ${\rm d}\tau_{{\rm obs}_m}$-stretched PSD for a simulated magnetized RG observed pole-on. \textsl{Middle panel:} ${\rm d}\pi_0$-LS periodogram for the simulated magnetized RG, using $\Delta\Pi_1=75.2$ sec and $\delta\nu_{\rm mag}(\nu_{\rm max})=0.40\, \mu$Hz. \textsl{Bottom panel:} LS periodogram of the ${\rm d}\pi_0$-stretched PSD for the simulated magnetized RG by using $\Delta\Pi_1=75.13$ sec and $\delta\nu_{\rm mag}(\nu_{\rm max})=0.38\, \mu$Hz (as calculated in Sect.~\ref{sec:fit}). In each panel, the black line indicates the true value of $\Delta\Pi_1$, and the dot-dashed blue line indicates the period associated with the maximum amplitude in the LS periodogram. Dashed grey curves are the mirrored central peak around the period associated with the maximum amplitude in the LS periodogram. Curves are thicker within the central peak for better visualization. {Hatched areas indicate the typical uncertainty $\delta\left(\Delta\Pi_1\right)_{\rm res}=0.6$ seconds on the measurement of $\Delta\Pi_1$ from LS-based methods according to the study of \cite{Vrard2016}}, centered around $\Delta\Pi_1=75.2$ sec. The skewness of the dominant peak calculated with Eq.~\ref{eq:skew} is indicated in the top left corner of each panel.}
    \label{fig:LSmag0}
\end{figure}

\noindent Using Eq.~\ref{eq:nuobsmag} (for the magnetized, non-rotating case) or Eq.~\ref{eq:nuobsrotmag} (for the magnetized, rotating case) for $m=0$ mixed modes, the equation describing the unperturbed frequencies of the modes is:% in the magnetized non-rotating case is:
\begin{equation}
    \nu^4-\nu_{{\rm obs}_0}\nu^3+\frac{\zeta A {\rm B}_0^2}{2}=0\, .
    \label{eq:numm}
\end{equation}

As we focus on $m=0$ modes, rotation does not play a role (for the slow rotation rates of red giants).
% and using Eq.~\ref{eq:nuobsrotmag} for 
% \begin{equation}
%     \nu^4(m)+\left(m\zeta\delta\nu_{\rm rot}-\nu_{\rm obs}\right)\nu^3(m)+m^*\zeta A=0
%     \label{eq:numm}
% \end{equation}
\noindent Equation~\ref{eq:numm} can be solved analytically to obtain $\nu=f_0(\nu_{{\rm obs}_0}),$ following Appendix~\ref{App:nu_mm}.
The measurement of $\Delta\Pi_1$ in the presence of internal magnetism therefore strongly depends on our ability to detect magnetically shifted $m=0$ dipolar modes in the PSD. \\%long as $i\approx 0\degree$.\\

To demonstrate the efficiency of the corrected $\pi$ stretching defined by Eq.~\ref{eq:dpi_mag_nu}, we stretch the PSD of a simulated magnetized star observed pole-on ($i=0\degree$) according to the $m=0$ magnetized stretch period $\pi_0$ defined as a function of the observed PSD:

\begin{equation}
        {\rm d}\pi_0 = -\frac{1}{\zeta}\frac{{\rm d}\nu_{{\rm obs}}}{f_0(\nu_{{\rm obs}})^2} \frac{1}{\displaystyle1 - \frac{3\zeta}{2} \frac{A{\rm B}_0^2}{f_0(\nu_{{\rm obs}})^4} } \, .
    \label{eq:dpi0}
\end{equation}

The top panel of Fig.~\ref{fig:LSmag0} shows the LS periodogram of the ${\rm d}\tau_{{\rm obs}}$-stretched PSD of a star seen pole-on (only $m=0$ components are visible in the PSD), with an axisymmetric magnetic field perturbing mixed-mode frequencies inside the radiative interior following Eq.~\ref{eq:nuobsmag}. The amplitude of the field is taken such as the magnetic perturbation at $\nu_{\rm max}$ (given by ${\displaystyle A {\rm B}_0^2}/{\displaystyle \nu_{\rm max}^3}$) is equal to $50$ times the frequency resolution, $\delta f_{\rm res}$, in the PSD. As $\delta f_{\rm res}\approx 0.008\, \mu$Hz for the four-year observations during the \textsl{Kepler} mission, the magnetic signature at $\nu_{\rm max}$ is chosen to be $0.4\, \mu$Hz.  %=50\delta f_{\rm res}$. 
% To compute the LS represented on the top panel, we use the prescription of the stretching of the PSD by the $\tau_{{\rm obs}_0}$ function from Eq.~\ref{eq:tau} as proposed by \cite{Vrard2016}. 
We observe that the measured period spacing (dotted-dashed blue line) is smaller than the true g-mode period spacing (black line), with an offset of about $1\%$ (as already discussed in Sect.~\ref{sec:mag}). % We therefore demonstrate that ignoring potential magnetic fields inside the radiative interior of the star might bias our estimates of the gravity mode period spacing. 
% Depending on the field strength, this could have large impacts on the measured internal rotation rate on the RGB by \cite{Gehan2018} which strongly relies on the accuracy of the $\Delta\Pi_1$ measure. 
We conclude that if we do not have any prior on the magnetic field amplitude inside the observed star, then we do not have any dependable way of obtaining an accurate value for $\Delta\Pi_1$.

For the middle panel of Fig.~\ref{fig:LSmag0}, we assume that we know the value of $\delta\nu_{\rm mag}$. We can then represent the LS periodogram of the ${\rm d}\pi_0$-stretched PSD (${\rm d}\pi_0$-LS periodogram) of the simulated magnetized RG according to Eq.~\ref{eq:dpi0}. We display a measurement of the period associated with the maximum power in the PSD (red dashed line) at $\Delta\Pi_{\rm LS}=75.19\sec$. The true period spacing used to simulate the star being $\Delta\Pi_1=75.20\sec$, it corresponds to an estimation error of the $\Delta\Pi_1$ parameter of about $0.01\%$ in the presence of magnetism through the use of $m=0$ modes alone. In Appendix~\ref{App:pi_1}, we provide the analogous results when considering $m=\pm1$ oscillation modes, which leads to a more complicated analysis in the case of rotation stars as $m=\pm1$ are excited with the same amplitude \citep{Gizon2003} and one cannot exist without the other. \\

\subsection{Effect of the inclination of the star on the accuracy of $\Delta\Pi_1$ with the ${\rm d}\pi_0$-LS method}

 The measurement of $\Delta\Pi_1$ with this "${\rm d}\pi_0$-LS" method is robust only for stars with relatively low inclination angles, as the amplitude of the $m=0$ modes must emerge above the noise level in the PSD. In the bottom panel of Fig.~\ref{fig:LSrotincl} we represent the ${\rm d}\pi_0$-LS periodogram of the simulated rotating and magnetized model (see the corresponding PSD in the bottom panel of Fig.~\ref{fig:PSDs}) observed with various inclination angles. When the star is observed pole-on, the ${\rm d}\pi_0$-stretching leads to a {nearly perfect} estimate of $\Delta\Pi_1$. As the inclination increases and the amplitude of the $m=\pm1$ components increase with respect to the $m=0$ component's amplitude, the ${\rm d}\pi_0$-stretching is not adapted to the data anymore, and the measurement of the period spacing becomes biased again. For the rest of this study, we set our focus on stars with $i \leq 30\degree$ to ensure that $\Delta\Pi_1$ is accurately estimated from the period associated with the maximum of the ${\rm d}\pi_0$-LS periodogram even when the star is rotating.

% As we don;t know a priori the value of $\delta\nu_{\rm mag}$ for observed stars, we base our detection on the measure of the skewness of the central peak in the LS. As mentioned in section~\rec{sec:skewness}, if a star is magnetized the use of the regular $\tau$ function leads to a positive skewness in the LS. When using the $\pi_0$ function, the skewness is brought closer to zero (meaning symmetrical normal distribution), as $S\approx0.08$ on the bottom panel of Fig.~~\ref{fig:LSmag0}.

% If the star is also rotating, Finally,
% \begin{equation}
%     {\rm d}\pi = -\frac{1}{\zeta}\frac{{\rm d}\nu_{{\rm obs}}_m}{\nu^2} \frac{1}{\displaystyle1 - 3m^*\zeta \frac{A{\rm B}_0^2}{\nu^4} } \, .
% \end{equation}

% The only difference with the case with magnetism only is therefore the determination of the mixed-mode unperturbed frequency $\nu$ from Eq.~\ref{eq:nuobsrotmag} instead of Eq.~\ref{eq:nuobsmag}

% \begin{equation}
%     \Delta\Pi_1 = -\int_{\nu_{{\rm obs}_m}} \frac{1}{\zeta}\frac{{\rm d}\nu_{\rm obs}}{\nu^2} \frac{1}{\displaystyle1 - 3m^*\zeta \frac{A {\rm B}_0^2}{\nu^4} } \, .
% \end{equation}

% \subsection{Extraction of the $m=0$ components from the complete PSD}

\subsection{Skewness of the dominant peak in the ${\rm d}\tau_{{\rm obs}_0}$-LS periodogram as a signature of internal magnetism}
\label{sec:skew}
On each panel of Fig.~\ref{fig:LSmag0} we mirror the dominant peak in the LS periodogram around the measured period spacing and represent it by the dashed grey curves. This visual trick allows to unveil the asymmetry of the dominant peak in the ${\rm d}\tau_{{\rm obs}}$-stretched LS periodogram (first panel), caused by the {dependency in frequency of the} magnetic effect. 
%By writing $p_{\rm peak}$ the array of length N periods corresponding to the peak with maximum amplitude in the LS periodogram and $A_{\rm peak}$ the corresponding amplitude in the LS periodogram, we measure the Fisher-Pearson coefficient of skewness of the peak distribution as a function of its mean value $\mu$ and standard deviation $\sigma$ through:
% \begin{equation}
%     S = \frac{\displaystyle \frac{1}{N}\left(\sum_{n=1}^N A_{{\rm peak}_n}(p_{{\rm peak}_n}-\mu)^3\right)}{\sigma^3}\, .
%     \label{eq:skew}
% \end{equation}%On the bottom panel the new p

% For the ${\rm d}\tau_{\rm obs}$-stretched PSD of the simulated star represented in the first panel of Fig.~\ref{fig:LSmag0}, the $S$ value is of about $0.39$. When using the appropriate ${\rm d}\pi_0$-stretching function via Eq.~\ref{eq:dpi0} the skewness is brought closer to zero (quasi-normal distribution), as $S\approx0.07$ in the middle panel of Fig.~~\ref{fig:LSmag0}.

Current interpretations of the oscillation spectra have ignored to search for skewness in the LS periodogram maxima. We prove here that non-zero skewness can be used as a detection method for internal magnetism {(see Appendix~\ref{app:skew} for more details)}. {Examining RGB stars more closely could reveal these small magnetic signatures.} 
%This aspect has never been investigated during the process of measuring internal rotation rates, and we might therefore have missed magnetic fields signatures in already well-studied stars in \cite{Vrard2016} and \cite{Gehan2018}. 
{However, other physical processes such as non-uniform rotation could lead to skewed distribution in the LS periodogram. Therefore, the skewness method should be used with care. In the following sections, we develop the main detection method for internal magnetic fields.}
% If an asymmetric signature in the dominant peak in the $\tau_{{\rm obs}_0}$-LS periodogram does now allow neither a direct measure of the magnetic shift of mixed-mode frequencies, nor a direct correction of the value of $\Delta\Pi_1$, it might be decisive in the process of flagging stars of potential interest for the search of internal magnetism. 

\subsection{Local measuresment of $\Delta\Pi_1$ in the PSD as a probe for internal magnetic fields}

We dig further into the ability to detect the presence of internal magnetism from LS-based measurements of $\Delta\Pi_1$, and develop a method to get a first estimate of the magnetic field amplitude without needing the very challenging fit of Eq.~\ref{eq:Mathis} to the PSD. The effect of an axisymmetric mixed poloidal and toroidal field buried in the radiative interior of RGB stars generates a characteristic signature on the different azimuthal $m$ components. All g-$m$ dipolar frequencies described by Eq.~\ref{eq:nuobsmag} \citep[see][]{Bugnet2021, Mathis2021} are shifted towards higher frequencies following the scaling law given by Eq.~\ref{eq:dnumag}. The shift of $m=\pm1$ modes is twice the shift of $m=0$ modes as represented in the third panel of Fig.~\ref{fig:PSDs}, but they all vary with the unperturbed frequency of the mode as $1/\nu^3$. The magnetic signature is thus much higher at lower frequencies in the PSD. The dependency of the magnetic shift with the mode frequency lies at the basis of the bias in the measurement of $\Delta\Pi_1$, shown in Fig.~\ref{fig:LSmag0}. In addition, the $\displaystyle {1}/{\nu^3}$ dependency changes the observed value $\Delta\tau_{\rm obs}$ along the observational frequency range considered here ($\nu \in $ [$\nu_{\rm max}-3\Delta\nu:\nu_{\rm max}+3\Delta\nu$]). We therefore investigate how local measurements of $\Delta\tau_{\rm obs}$ for various frequency ranges might allow to detect the presence of buried magnetic fields.\\

\begin{figure*}[t]
    \centering
    \includegraphics[width=1\textwidth]{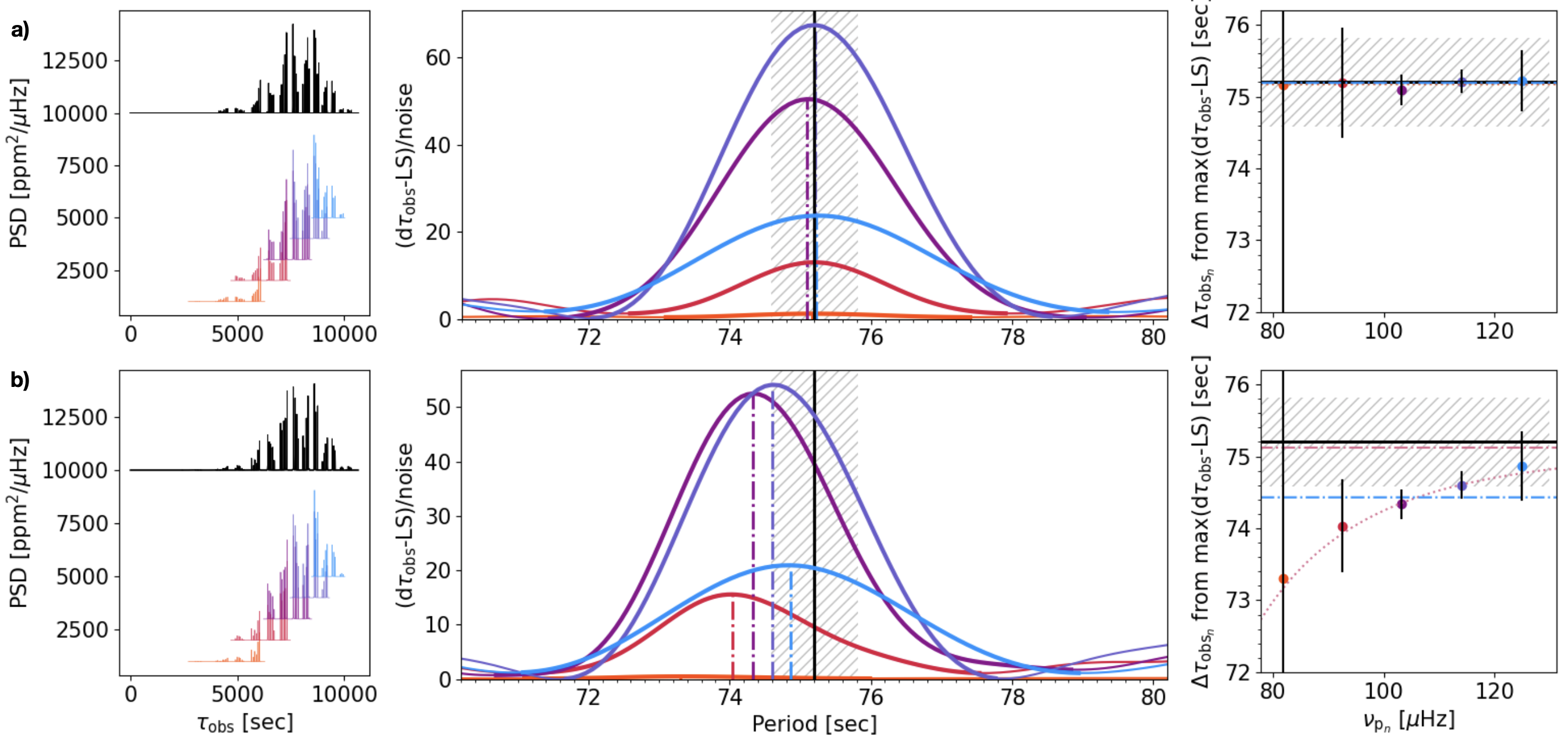}
    \caption{Method to retrieve $\Delta\Pi_1$. \textbf{a)} Non-rotating, non-magnetized model as in the first panel of Fig,~\ref{fig:PSDs}. \textbf{b)} Non-rotating, magnetized model as in the third panel of Fig,~\ref{fig:PSDs}. In each case, left panels represent the stretched PSD according to the ${\rm d}\tau_{\rm obs}$ variable (Eq.~\ref{eq:deltatau_obs}). Subsamples centered around the p-$m$ modes with a width of $2-\Delta\nu$ are extracted and plotted in different colors below one another. Vertical positions are chosen so that data do not overlap too much, they are therefore arbitrary. Middle panels show a zoom around the dominant peak in the normalized ${\rm d}\tau_{\rm obs}$-LS periodogram associated with the different frequency ranges color-coded on left panels. Curves are thicker within the central peak for better visualization. Vertical dot-dashed lines indicate the measurement of $\Delta\tau_{{\rm obs}_n}$ on the various frequency ranges. {Hatched areas indicate the typical uncertainty $\delta\left(\Delta\Pi_1\right)_{\rm res}=0.6$ seconds on the measurement of $\Delta\Pi_1$ from LS-based methods according to the study of \cite{Vrard2016}}, centered around $\Delta\Pi_1=75.2$ sec (same in the right panels). In right panels such {local} measurements of $\Delta\tau_{{\rm obs}_n}$ are transferred as a function of the central p-$m$ frequencies. The black line indicates the true value of $\Delta\Pi_1$ and the blue dash-dotted line indicates $\Delta\tau_{\rm obs}$. In the bottom panel, the dotted pink curve is the result of the Bayesian fit of the $\Delta\Pi_1$-$\nu_{{\rm p}_m}$ law according to Eq.~\ref{eq:function}, and the pink dash-dotted line indicates the measurement of $\Delta\Pi_1$ obtained from the fitting process (see Table~\ref{tab:fit}).}%\textsl{Bottom row:} Magnetized star perfectly corrected from the effect of magnetism.}
    \label{fig:dpi_n}
\end{figure*}

To do so, we create subsamples of the simulated model's ${\rm d}\tau_{\rm obs}$-stretched PSD on ($2\Delta\nu$)-wide intervals around each of the $\nu_{{\rm p}_n}$ modes as represented in the left panels of Fig.~\ref{fig:dpi_n}. We compute the ${\rm d}\tau_{\rm obs}$-LS periodogram for each sub-PSD (see middle panels of in Fig.~\ref{fig:dpi_n}) and we measure the period associated with their maximum power. In the right panel, we indicate these $\Delta\tau_{{\rm obs}_n}$ values as a function of $\nu_{{\rm p}_n}$. The row a) in Fig.~\ref{fig:dpi_n} represents a non-magnetized star observed pole-on. None of the $m=0$ visible modes are affected by any internal processes, therefore, all the measurements of $\Delta\tau_{{\rm obs}_n}$ are consistent with the value of $\Delta\Pi_1$ as represented in the top right panel. The b) star is magnetized with $\delta\nu_{\rm mag,g}(\nu_{\rm max})=0.4\, \mu$Hz, and observed pole-on. As expected from the $1/\nu^3$ dependency of the frequency shift, the period spacing $\Delta\tau_{{\rm obs}_n}$ measured from the ${\rm d}\tau_{\rm obs}$-LS periodogram increases and gets closer to the known $\Delta\Pi_1$ value as the $\nu_{{\rm p}_n}$ frequency increases (see bottom-right panel). %This is expected as the effect of the magnetic field on the frequencies of the $g-m$ modes increases at low frequency. 

When using Eq.~\ref{eq:uncertainty_res} to account for measurement errors from the LS method as done by \cite{Vrard2016}, uncertainties on all $\Delta\tau_{{\rm obs}_n}$ measurements should be equal. However, the signal-to-noise ratio (S/N) decreases from $\nu_{\rm max}$ towards the edges of the $[\nu_{\rm max}-3\Delta\nu:\nu_{\rm max}+3\Delta\nu$] interval. The measurement of the periodicity in the stretched spectrum should therefore take into account that the white noise in the PSD will not pollute the signal on every interval in the same way. The impact of the amplitude of the signal in the stretched PSD on the amplitude in the LS periodogram is visualized in middle panels of Fig.~\ref{fig:dpi_n}. We therefore use the oversampling resolution from Eq.~\ref{eq:uncertainty_over}, which is anti-correlated to the peak amplitude. The measurement of the oversampling uncertainty requires a normalization to the noise level in the LS periodogram. Here, we chose not to include any noise in our artificial spectra as the distribution and amplitude of noise might vary a lot in real observations. By drawing a few pure $\chi^2_2$ white noise PSDs and performing their LS periodogram, we estimate the typical contribution of the photon noise corresponding to a high S/N in the PSD and normalize the result. We then obtain a normalized LS periodogram as represented in Fig. \ref{fig:dpi_n}, with the normalized amplitude of the highest peak of a few dozens, representative of a real RGB asteroseismic data \citep[as discussed in][]{Vrard2016}. The orange peak corresponding to a frequency range in the PSD with modes of very small amplitude have an amplitude in the LS periodogram considered too low with respect to the noise level \citep[i.e., A<13,][]{Vrard2016} for the detection to be precise, which explains the large uncertainty reported in the right panels when measuring $\Delta\tau_{{\rm obs}_n}$ at low frequency.\\
% As discussed in \cite{Vrard2016}, typical dominant peak amplitudes $A$ are largely above the threshold allowing much smaller amplitudes than the resolution in the LS. This justifies the oversampling we performed on the LS (TO MOVE SOONER). 

We consider this $\delta\left(\Delta\Pi_{1_{\rm over}}\right)$ uncertainty as the smallest possible uncertainty. Indeed, very small uncertainties can be reached only when the gravity offset $\epsilon_{\rm g}$, intervening in the pure g-mode pattern, is perfectly known. As this is not strictly the case in real data, we also perform the following study by taking into account the resolution uncertainty defined by Eq.~\ref{eq:uncertainty_res}.

\subsection{Estimation of the magnetic splitting from ${\rm d}\tau_{{\rm obs}_0}$-stretched PSD subsamples}
\label{sec:fit}
We use the trend in the $\Delta\tau_{{\rm obs}_n}$ - $\nu_{{\rm p}_n}$ diagram to correct the $\Delta\Pi_1$ measurement for magnetized stars observed pole-on and to constrain the value of the magnetic splitting.
{Cutting the spectra into smaller frequency intervals allows to approximate Eq.~\ref{eq:dpi_mag_nuobs}. By considering $\nu\approx\nu_{{\rm p}_n}$ on each small frequency intervals:
\begin{equation}
\Delta\tau_{{\rm obs}_{n}} \approx \mathcal{C}\left({\nu_{{\rm p_n}}}, \delta\nu_{\rm mag,g}\right)\times \Delta\Pi_1 \, ,
\label{eq:obsn}
 \end{equation}
 \noindent with
 \begin{equation}
     \mathcal{C}\left({\nu_{{\rm p_n}}}, \delta\nu_{\rm mag,g}\right) =\frac{\displaystyle 1 - \frac{3}{2}\zeta_{\rm g}\left({\nu_{{\rm p}_n}}\right) \frac{\displaystyle \delta\nu_{\rm mag,g}}{\displaystyle \nu_{{\rm p}_n}} }{\displaystyle\left(1+\frac{1}{2}\zeta_{\rm g}\left({\nu_{{\rm p}_n}}\right)\frac{\delta\nu_{\rm mag,g}}{\nu_{{\rm p}_n}}\right)^2}\,.
 \end{equation}
{With this approximation, we consciously underestimate the effect of the magnetic field on the frequencies of the modes and we therefore underestimate the value of $\Delta\Pi_1$.} However, this loss of accuracy is counterbalanced by the possibility of performing a simple fit of the $\left(\Delta\Pi_1, {\delta\nu_{\rm mag,g}}\right)$ parameters from the $\Delta\tau_{{\rm obs}_n}$-$\nu_{{\rm p}_n}$ data represented of the right panels.} We thus fit the value of the magnetic parameter $\delta\nu_{\rm mag,g}$ along with $\Delta\Pi_1$ to adjust the following function derived from Eq.~\ref{eq:obsn} in the $\Delta\tau_{{\rm obs}_n}$-$\nu_{{\rm p}_n}$ diagram:
\begin{equation}
    \mathcal{F}\left(\Delta\Pi_1, {\delta\nu_{\rm mag,g}}\right): \nu_{{\rm p}_n} \rightarrow \mathcal{C}\left({\nu_{{\rm p_n}}}, \delta\nu_{\rm mag,g}\right)\times \Delta\Pi_1 \, .
    \label{eq:function}
\end{equation}
% \textcolor{red}{\begin{equation}
%     \mathcal{F}\left(\Delta\Pi_1, {A{\rm B}_0^2}\right): \nu_{{\rm p}_n} \rightarrow {\Delta\Pi_{{1}}}{\frac{\left(1-\displaystyle \frac{3}{2}\zeta_{\rm g}\left({\nu_{{\rm p}_n}^4}\right)\frac{A{\rm B}_0^2}{\nu_{{\rm p}_n}^4}\right)}{\left(1+\displaystyle\frac{ 1}{2}\zeta_{\rm g}\left({\nu_{{\rm p}_n}^4}\right)  \frac{A{\rm B}_0^2}{\nu_{{\rm p}_n}^4}\right)^2}}\, ,
% \end{equation}CAN WE DO BETTER???}

% \noindent with $\zeta_{\rm max}$ the asymptotic expression of the $\zeta$ function associated modes domiated by the gravity component:
% \begin{equation}
%     \zeta_{\rm max} = \left(1+\frac{q}{\mathcal{N}}\right)^{-1}.
% \end{equation}
First, we perform a non-linear least-squares fit to the data points in the $\Delta\tau_{{\rm obs}_n}$-$\nu_{{\rm p}_n}$ diagram, from which we obtain the following values: $\Delta\Pi_1=75.11 \,{\rm sec}\, {\rm and}\, {\delta\nu_{\rm mag,g}}=0.39 \,\mu{\rm Hz}$. Then, we use these values as priors for a Bayesian fitting on the same data points, by using the \texttt{emcee} python ensemble \citep[see Appendix~\ref{app:fit} for the details of the fit,][]{ForemanMackey2019}. The result of the Bayesian fit is represented by the dash-dotted pink curve in the lower right panel of Fig.~\ref{fig:dpi_n}. With this diagram, we offer a way of detecting the presence of magnetic fields affecting the frequency of g-$m$ modes. The continuous pink line indicates the value of $\Delta\Pi_{1_{\rm LS}}$ as estimated from the fit to the data with $\delta\left(\Delta\Pi_{1_{\rm over}}\right)$ uncertainties (see Table~\ref{tab:fit}). As discussed, it is lower than the true value of $\Delta\Pi_1$ due to the $\nu\approx \nu_{{\rm p}_n}$ approximation. The error of $\Delta\Pi_1$ for this specific simulated model is $\left|\Delta\Pi_1-\Delta\Pi_{1_{\rm LS}}\right|/\Delta\Pi< 0.01\%$, within the typical error associated with classical methods (from $\delta\left(\Delta\Pi_{1_{\rm res}}\right)$) as represented by the hatched area in the right panels of Fig.~\ref{fig:dpi_n}. Here, $\delta\nu_{\rm mag,g}$ is underestimated by about $6\%$.
%We here consider the $\delta\left(\Delta\Pi_{1_{\rm over}}\right)$ uncertainty as the smallest possible uncertainty, associated with a perfect estimate of $\epsilon_{\rm g}$ (see Sec.~\ref{}). %Indeed, very small uncertainties can be reached only when the gravity offset $\epsilon_{\rm g}$ intervening in the pure gravity-mode pattern is perfectly known, which will not be the case in real data. 
For consistency, we also perform the Bayesian fit by taking into account the resolution uncertainty defined by Eq.~\ref{eq:uncertainty_res}, and the corresponding results are reported in the last row of Table~\ref{tab:fit}. The magnetic effect is underestimated by about $12\%$ in this case, and the error on $\Delta\Pi_1$ is slightly larger than when using $\delta\left(\Delta\Pi_{1_{\rm over}}\right)$. This larger departure from the true value of each parameter is expected, as a lot of weight is given to less significant measurements when using $\delta\left(\Delta\Pi_{1_{\rm res}}\right)$.

% Applying this methodology to real stars observed quasi pole-on ($i\leq 30\degree$) is fairly simple as opposed to the multi-parameters fit on the PSD that has been the only option so far. In addition to being a powerful tool for the detection of magnetized stars, this method also allows to get a reasonable estimate of the magnetic frequency splitting inside magnetized stars.\\

\tabcolsep=2pt 
\begin{table*}
\centering

\caption{Comparison of the performances of the $\tau_{\rm obs}$ method developed by \cite{Vrard2016} and the $\pi_0$ fitting method from Sec.~\ref{sec:fit} in the case of a non-magnetized \textbf{a)} and a magnetized \textbf{b)} star, depending on the prescription chosen for uncertainties of $\Delta\tau_{\rm obs}$ measurements.}
\begin{tabular}{cll|lr|lr|r}
 & & True $\delta\nu_{\rm mag,g}(\nu_{\rm max})$ &  $\delta\nu_{\rm mag,g}(\nu_{\rm max})$ & Uncertainty & $\Delta\Pi_{1_{\rm LS}}$  & Uncertainty & p-value \\
 & & [$\delta f_{\rm res}$] & [$ \delta f_{\rm res}$] & [$ \delta f_{\rm res}$] &  [sec] &  [sec] &\\
 &&&&&&&\\

\hline
%  &&&&\\
 \textbf{a)}&$\tau_{\rm obs}$ method&0 &  \o&  \o& 75.19 & $\delta\left(\Delta\Pi_{1_{\rm res}}\right) = 0.61$ &\o\\
  & $\pi_0$ fitting method, $\delta\left(\Delta\Pi_{1_{\rm over}}\right)$&0 & \o& \o& 75.17 & 0.11 &0.13\\
%   $\pi_0$ local fitting method, $\delta\left(\Delta\Pi_{1_{\rm over}}\right)\times 10&0 & \o& \o& 75.16 & 0.31 &0.60\\
 & $\pi_0$ local fitting method, $\delta\left(\Delta\Pi_{1_{\rm res}}\right)$&0 & \o& \o& 75.18 & 0.27 & 1\\
%  &&&&\\
 &&&&&&&\\
 \hline

% $\nu_{\rm max}*f_{\rm res}*50 \approx 508,200$ & 75.05 & 438,900\\
 \textbf{b)}& $\tau_{\rm obs}$ method&$50$& \o& \o & 74.43 & $\delta\left(\Delta\Pi_{1_{\rm res}}\right) = 0.61$ & \o\\
 & $\pi_0$ local fitting method, $\delta\left(\Delta\Pi_{1_{\rm over}}\right)$&$50$&  47& 32 & 75.13 & 0.45 & <0.001\\
%  $\pi_0$ local fitting method, $\delta\left(\Delta\Pi_{1_{\rm over}}\right)\times 10&$50$&  49& 90 & 75.13 & 1.3 & 0.06\\
 
 & $\pi_0$ fitting method, $\delta\left(\Delta\Pi_{1_{\rm res}}\right)$&$50$&  44& 22 & 75.12 & 0.53 & <0.001\\
\end{tabular}

\label{tab:fit}
\end{table*}

\subsection{Confidence in the magnetic detection}

Even if the star is not magnetized, it is always possible to fit the magnetized function, $\mathcal{F,}$ to the data, and to extract a small magnetic splitting value. However, such a measurement is probably not significantly better than a fit with  $B_0=0$, which means that it is necessary to assess the robustness of any magnetic detection from the Bayesian fit. To do so, we compare the result of the magnetized fit (Eq.~\ref{eq:function}) and those from the fit of a non-magnetized "nested" model $\mathcal{F}_0$ defined as:

\begin{equation}
    \mathcal{F}_0\left(\Delta\Pi_1\right):\nu_{{\rm p}_n} \rightarrow \Delta\Pi_1\, .
    \label{eq:function_nomag}
\end{equation}

\noindent To do so, we perform a likelihood-ratio chi-squared test to approve or reject the null hypothesis ${\rm H}_0$: \textbfediting{"{The full model and the nested model fit the data equally well. Thus, the nested model should be used".}} We calculate the p-value for the null model using the $\chi^2_{{\rm df}=1}$ cumulative distribution function:
%  survival function of $\lambda _{\text{LR}}$ with a $chi^2_1$ distribution. %http://rnowling.github.io/machine/learning/2017/10/07/likelihood-ratio-test.html
\begin{equation}
    {\rm p} = P[\chi^2_{{\rm df}=1}>\lambda _{\text{LR}}]\, ,
    \label{eq:pvalue}
\end{equation}
with $\lambda _{\text{LR}}$ the ratio of the log likelihoods:
\begin{equation}
    {\displaystyle \lambda _{\text{LR}}= 2\left[~\log\mathcal{L} ({\mathcal{F}})-\log\mathcal{L} ({{{\mathcal{F}_0} }})~\right]}\, .
\end{equation}

The ${\rm p}$-value defined in Eq.~\ref{eq:pvalue} is the probability that the null hypothesis is true.
A ${\rm p}$-value lower than 0.05 means that we can reject the null hypothesis and consider the star to be magnetized with a level of confidence of $95\%$. In the case of the artificial magnetized star studied in Sect.~\ref{sec:fit}, the $p$-value is $<0.001$. We, therefore, reject the null hypothesis and confirm the detection of magnetism from the LS method. In the case of the non-magnetized star, the ${\rm p}$-value is above $0.05,$ therefore, the null hypothesis should not be rejected and the nested non-magnetic model ought to be considered. We confirm the non-detection of magnetism in the artificial non-magnetized star.
% See Fig.~\ref{fig:dpi_n}

% \begin{table*}
% \centering

% \caption{Comparison of the performances of the $\tau_{\rm obs}$ method developed by \cite{Vrard2016} and the $\pi_0$ fitting method from Sec.~\ref{sec:fit} in the case of a non-magnetized and a magnetized star.}
% \begin{tabular}{l|cc|cc}
%  & $\delta\nu_{\rm mag,g}(\nu_{\rm max}) &  Error on $\delta\nu_{\rm mag,g}(\nu_{\rm max})$ & $\Delta\Pi_{1_{\rm LS}}$  & Uncertainty on the measure \\
%   & [$\delta f_{\rm res}$] &   [$ \delta f_{\rm res}$] &  [sec] &  \\
%  &&&&\\

% \hline
% %  &&&&\\
% Non-magnetized, $\tau_{\rm obs}$ method&0 &  /o& 75.19 & 0.00\%\\
% Non-magnetized, $\pi_0$ local fitting method&0 &  \sim 0& 75.17 & 0.04\%\\
% %  &&&&\\
%  \hline

% % $\nu_{\rm max}*f_{\rm res}*50 \approx 508,200$ & 75.05 & 438,900\\
% Magnetized, $\tau_{\rm obs}$ method&$50$&  /o & 74.43 & 1.02\% \\
% Magnetized, $\pi_0$ local fitting method&$50$&  \sim -7& 75.05 & 0.20\%\\
% \end{tabular}

% \label{tab:fit}
% \end{table*}
% \subsection{Effect of the photon noise}

\section{Results and discussion}
\label{sec:result}

The bottom panel of Fig.~\ref{fig:LSmag0} presents the ${\rm d}\pi_0$-LS periodogram computed when using the output of the fitting process: $\Delta\Pi_1=75.13\,{\rm sec}$ and ${\delta\nu_{\rm mag,g}}=0.38 \,\mu{\rm Hz}$. As demonstrated by the comparison of top and bottom panels, we are able to largely reduce the error committed from the ${\rm d}\tau_{\rm obs}$ method on the measurement of $\Delta\Pi_1$ when the star is magnetized (as also reported on Table~\ref{tab:fit}). The new measurement of $\Delta\Pi_{1_{\rm LS}}$ from this "observational" ${\rm d}\pi_0$-LS periodogram is $75.15$ sec (typical $\delta\left(\Delta\Pi_{\rm res}\right) =0.6 $ sec), coherent with the value resulting from the fit, and close to the true value of 75.2 seconds. The skewness of the dominant peak has also been largely reduced from $0.39$ to $0.06$ (see Appendix~\ref{app:skew} for more details about the skewness of the dominant peak). %A very small error remains on the estimate of $\Delta\Pi_1$ and $\delta\nu_{\rm mag,g}$ as the middle and the last panels of Fig.~\ref{fig:LSmag0} are not identical. %The $\pi_0$-LS fitting method should therefore be used to flag magnetized red giants, and as a prior to a complete fit of the modes in the PSD to avoid local minima in the fitting process.

\subsection{Detectability of the magnetic field through local $\Delta\Pi_1$ measurements}

The fitting method presented in Sec.~\ref{sec:fit} relies on local measurements of $\Delta\tau_{{\rm obs}}$, which have intrinsic uncertainties given by Eq.~\ref{eq:uncertainty_over}. The $1/\nu^3$ signature of the magnetic frequency splitting might not emerge significantly from the method described in Sect.~\ref{sec:measure_dpi}, with a ${\rm p}$-value higher than about $0.05$. This might happen when the ratio $A{\rm B}_0^2/\nu^3$ leads to magnetic signature on the order of only a few frequency resolutions or less when the signal to noise ratio in the spectrum leads to large $\delta\left(\Delta\nu_{\rm over}\right)$ uncertainties or if the amplitude of the magnetized dipolar modes is also suppressed by the internal magnetic field  \citep[e.g.,][]{Fuller2015, Loi2017}. To get more confidence in an eventual magnetic detection, we aimed to select stars combining a significant skewness in the dominant peak in the ${\rm d}\tau_{\rm obs}$-LS periodogram with a strong variation of the $\Delta\tau_{{\rm obs}_n}$ measurements with the frequency leading to a low ${\rm p}$-value during the fitting process.

% \subsection{Estimation of $\delta\nu_{rm mag}$ for xxx red giants observed by \kepler}
% \subsection{Forward modelling to extract B_0}
\subsection{Confusion with other sources of frequency shifts}

As extensively discussed in \cite{Bugnet2021}, known sources of frequency perturbation do not produce signatures that resemble magnetic effects:\\

\textbf{Latitudinal differential rotation} produces perturbations that are typically smaller than the frequency resolution \citep{Deheuvels2017}. Therefore, they are neglected in this study.\\

\textbf{Second-order rotational effects} As the star rotates, second-order and higher order asymmetric rotational perturbations of the centrifugal and Coriolis accelerations affect the perfectly-symmetric first-order rotational frequency pattern \citep{Dziembowski1992, Suarez2006}. \cite{Deheuvels2017} demonstrated that the observed rotation rates on the RGB \citep{Gehan2018} are too low for second-order rotational effects to significantly modify the frequency pattern of RGB stars.\\

\textbf{Structural glitches}. Sharp structural variations located in the deep layers of stars (hereafter, "glitches") are known to modify the frequency of mixed modes \citep{Cunha2015a}. As a result, the observed period spacing of mixed modes can be affected by the structure of the star and strongly vary with the frequency. The study of \cite{Cunha2015a}, which focuses on the presence of glitches on the RGB shows that such processes arise mostly during very short events such as the RGB luminosity bump, at the early phases of helium core burning, as well as at the asymptotic giant branch bump. {An asymmetric rotational triplet due to buoyancy glitches is then expected, caused by strong chemical gradients generated by the first dredge-up and left behind by the retreating envelope \citep{Cunha2015a, Cunha2019, Jiang2020}. \cite{Mosser2018} demonstrate that KIC 3216736 is the only RG among the 200 studied that presents buoyancy glitches. As a consequence, glitches are not a concern for the very large fraction of observed RG and this effect is neglected in the rest of our study.} We point out that the search for magnetic fields according to the method described in this work should be carefully carried out on stars away from these evolutionary stages to avoid false detections. \\

\textbf{Near-degeneracy effects} result from the combination of rotation and mode mixing \citep{Dziembowski1992, Suarez2006}. We refer to the complete study of \cite{Deheuvels2017} for the theoretical development of near-degeneracy effects on the asymmetry of rotational multiplets. Thsee authors demonstrated that near-degeneracy effects should be taken into account when studying the frequency pattern along the RGB. However, the near degeneracy effect does not produce a pattern varying in $1/\nu^3$ as magnetism does. For this reason, near-degeneracy effects should, in most cases, not be confused with a magnetic detection. We also emphasize the fact that near-degeneracy effects produce larger asymmetry measurements for larger order $\ell,$ as opposed to magnetic effects \citep{Bugnet2021}, which adds a powerful diagnostic to disentangle the two mechanisms. {The full comparison of near degeneracy effects and magnetic signature is the object of the study by \citep{Ong2022}.}\\

\subsection{The $\Delta\Pi_1$-$\Delta\nu$ diagram: Impact of the mass}

Representing the observed $\Delta\Pi_1$ values versus asteroseismic $\Delta\nu$ is a powerful visualization tool to disentangle stars from the RGB and from the red clump, as $\Delta\Pi_1$ is strongly modified during the transition. When focusing on the RGB, \cite{Vrard2016} noticed a dependency of the $\Delta\Pi_1$ value on stellar mass. Indeed, an intermediate-mass RGB exhibits a greater $\Delta\nu$ value than a low- mass star at fixed $\Delta\Pi_1$. It has been interpreted as the result of the difference in the density profile as observed from simulations \citep{Stello2013a}, because intermediate-mass stars are more dense than low-mass stars at given stellar core properties (fixed $\Delta\Pi_1$).\\

It is, however, also possible to reverse the scenario: intermediate-mass RG with ($1.3{\rm M}_\odot<M_{\star}<1.6{\rm M}_\odot$) present $\Delta\Pi_1$ values that are slightly lower than low-mass RG ($0.8{\rm M}_\odot<M_{\star}<1.3{\rm M}_\odot$) at fixed $\Delta\nu$. As demonstrated in our study, non-detected internal stable magnetic fields might have artificially biased the measurements of the $\Delta\Pi_1$ of intermediate-mass stars towards lower values. The mass dependency observed by \cite{Vrard2016} on the RGB could therefore also fully or partially artificially result from (undetected) magnetic fields inside intermediate-mass stars. The $\Delta\Pi_1$-$\Delta\nu$ diagram might therefore be used to flag candidates for a search for internal magnetic fields.

% cf \cite{Loi2021}: axi triplet dominates the frequency pattern.

% \subsection{Estimating radial magnetic field amplitudes from magnetic splittings}
\section{Conclusion and perspectives}
\label{sec:ccl}

In this work, we investigate how magnetic fields impact the measurement of the gravity mode (g-mode) period spacing from asteroseismic data of stars on the RGB and how such a measurement might allow us to detect the presence of buried magnetic fields inside the radiative interior of evolved stars. We created artificial RG power spectral densities (PSD) from mixed-mode frequency patterns \citep{Mosser2015} and we account for rotational \citep[e.g.,][]{Gizon2003}, and magnetic perturbations on mode eigenfrequencies \citep{Bugnet2021, Mathis2021}. By performing a Lomb-Scargle periodogram on the PSD and extracting the period associated with the peak of maximum amplitude, as done by \cite{Vrard2016}, we evaluate the effect of magnetic fields on the estimation of $\Delta\Pi_1$.\\

We demonstrate that g-mode period-spacing estimates, directly measured from asteroseismic data following \cite{Vrard2016}, might be biased towards lower values due to the presence of stable magnetic fields buried inside RGB stars \citep{Bugnet2021}. The shift of $\Delta\Pi_1$ measurements due to magnetism is small enough ($\sim 1\%$) for magnetic stars to blend in with non-magnetic stars when performing automatic measurements of period spacing. Magnetic stars might therefore be hidden among stars already studied in the past. In addition to $\Delta\Pi_1$ being shifted by the presence of magnetic fields inside the radiative interior of RGs, we show that the shape of the dominant peak in the LS periodogram that facilitates this measurement is also affected. A large skewness of the dominant peak in the LS periodogram of the ${\rm d}\tau_{\rm obs}$-stretched PSD is a first step to flag candidate stars for an optimal search of internal magnetic field signatures. This study relies on a first-order perturbative analysis of the magnetic field effects on mixed-mode frequencies performed in \cite{Bugnet2021}. The critical magnetic-field amplitude above which the perturbative analysis is not valid has been estimated by \cite{Bugnet2021}: the expected magnetic fields amplitudes inside stars on the RGB are of about 1000 times lower than this critical field. We therefore confirm that the perturbative analysis presented here is valid for stars evolving on the red giant branch.%we conclude that Results presented here are therefore valid in the perturbative regime. Which corresponds to magnetic field amplitudes bellow $\sim 1-10$ MG for stars with $\nu_{\rm max}\g 100\, \mu$Hz during the RGB \citep{Bugnet2021}. This method is thus suitable for young stars on the red giant branch that are expected to host fields\\

In addition, due to the frequency dependency of the magnetic effect on mixed-mode frequencies \citep[see][]{Mathis2021}, the $\Delta\Pi_1$ measured from various nested frequency intervals from a same PSD are different. Indeed, low-frequency $\Delta\Pi_1$ values are much lower than $\Delta\Pi_1$ computed from high-frequency intervals, as the effect of magnetism is the strongest on low-frequency g-$m$ modes. We therefore provide the correct stretching function for magnetized stars, to account for magnetic effects when measuring the ${\rm g}$-mode period spacing. We demonstrate that the variation of $\Delta\Pi_1$ measurements with frequency allows to estimate the magnetic splitting as defined in \cite{Bugnet2021} and \cite{Mathis2021} as well as to correct for the measured value of $\Delta\Pi_1$.\\

We combine theoretical prescriptions with realistic simulations of asteroseismic observations to construct the first magnetic detection method {based on the period spacing of mixed modes}. This study opens the way for an extensive search of magnetic fields inside evolved stars with solar-like oscillations. Due to the known effect of magnetic fields on mixed-mode frequencies from \cite{Bugnet2021}, the magnetic signature on local measurements of the period spacing of g-modes investigated here should not get mistaken for another known physical process also leading to frequency shifts (e.g., latitudinal differential rotation, centrifugal effects, glitches, near-degeneracy effects). Such a detection of magnetic-field signatures from the measurement of the period spacing of g-modes can then be inverted to get an estimate of the radial magnetic-field strength inside RGs following \cite{Mathis2021}, as it is already possible for the internal rotation \citep[e.g.,][]{Deheuvels2017, Deheuvels2020}. Our methodology is well-suited for stars observed with low-inclination angles or observed pole-on, as it is based on azimuthal order $m=0$ components to avoid additional signatures in the LS periodogram resulting from the distinct periodicity of $m=\pm 1$ modes. {As long as the magnetic-field amplitude remains moderate, the method presented here can be applied even if the field is inclined inside the star (see Appendix~\ref{App:inclined})}. From the statistics on the amplitudes of oscillations on the RGB \citep[e.g.][]{Stello2016a}, intermediate-mass stars seems more likely to host strong magnetic fields than low-mass stars during the RGB. There are a few stars in the study of \cite{Gehan2021} that are observed with low-inclination angles, with a mass above $1.3$ solar masses, and that present a rather low value of $\Delta\Pi_1$ from the study of \cite{Vrard2016}. The detection of magnetic fields inside these stars is out of the scope of this paper and its focus on methodology, but we are able to identify these stars as the best sample for the search of buried magnetic fields. The observational search is made possible thanks to the theoretical progress made in this study. Such a detection would be game-changing for improving the understanding of all stars of low and intermediate mass, as they are described as non-magnetic bodies in current stellar evolution models. The presence of strong internal magnetic fields inside the radiative interior of stars along their evolution would modify stellar age estimates and would thus affect many astrophysical areas -- from planet habitability studies to galacto-archeology \citep{Rauer2014, Chaplin2020}.

\begin{acknowledgements}
The author thanks S. Mathis, R.A. García, S. Mathur, C. Aerts, E. Corsaro, J. Fuller and M. Cantiello for very useful discussions. {The author thanks the referee for their valuable time in reviewing the manuscript and providing suggestions for improvement.} This research was supported in part by the National Science Foundation under Grant No. NSF PHY-1748958. {This research made use of Astropy,\footnote{http://www.astropy.org} a community-developed core Python package for Astronomy \citep{Robitaille2013, Price-Whelan2018}.}
    % \bld{We thank the referee for very useful and detailed comments that allow to improve the quality of our study and the article.} L. Bugnet, V. Prat, S. Mathis, A.A stoul, and K. Augustson acknowledge support from the European Research Council through ERC grant SPIRE 647383. All CEA members acknowledge support from GOLF and PLATO CNES grants of the Astrophysics Division at CEA. S. Mathur acknowledges support by the Ramon y Cajal fellowship number RYC-2015-17697. L. Amard acknowledges funding from the European Research Council (grant agreement No. 682393 AWESoMeStars). We made great use of the megyr python package for interfacing MESA and GYRE codes.In honour of our dear friend and colleague Michael J. Thompson. À la plus belle étoile de mon ciel.
\end{acknowledgements}

% \nocite{Mathis2021}
\bibliographystyle{aa}  
\bibliography{References}

%\newpage
\begin{appendix}
\section{Mixed-mode height and linewidth}
\label{App:linewidthheight}

% \subsubsection{Mixed-mode height and linewidth}

To create artificial RG power densities, we use the following prescriptions for mixed-mode amplitudes and linewidths.
The linewidth of each pure pressure mode is approximated, following {the empirical fit of mode linewidths from the MS to the RGB by} \cite{Corsaro2012}, as:
\begin{equation}
    \Gamma_{\rm p} \approx 1.39 \exp{\left(\displaystyle \frac{T_{\rm eff}-5777}{604}\right)}\, .
\end{equation}
The height of a resolved acoustic dipolar mode $H_{1_{\rm p}}$ observed by \kepler{} can be approximated by \citep{Mosser2012, Mosser2017}:
\begin{equation}
H_{1_{\rm p}}^2 = \frac{H_{0_{\rm p}^2}} {1.54}\, .%\left(1-\zeta\right)
\end{equation}
%\noindent with the $1.54$ geometrical factor adapted for the visibility of the $\ell=1$ modes observed by \kepler{}  .
As modes are often not resolved in the PSD, a dilution factor is taken into account \citep{Mosser2018} and $H_{1_{\rm p}}$ becomes:
\begin{equation}
    H_{1_{\rm p}} = \frac{\pi}{2}\frac{\Gamma}{\delta {f}_{\rm res}}\frac{H_{0_{\rm p}}} {\sqrt{1.54}}
,\end{equation}
\noindent
with $\delta {f}_{\rm res}$ as the frequency resolution in the PSD ($\delta {f}_{\rm res}\approx 0.008\, \mu$Hz for four-year observations with the \textsl{Kepler} satellite simulated in this work). It is not only the frequencies, but also the amplitudes ($A=\frac{\displaystyle \pi}{2}\Gamma H$) and linewidths that are modified via the coupling with g-modes \citep{Benomar2014, Belkacem2015}, following:
\begin{equation}
    \Gamma(\nu)=\Gamma_{\rm p}\left(1-\zeta(\nu)\right)
\end{equation}
\noindent and
\begin{equation}
    A^2(\nu)=A_{\rm p}^2\left(1-\zeta\right)\, .
\end{equation}
In addition, amplitudes in the PSD are modulated by the inclination angle of the star compared to the line of sight $i$, following \citep[e.g.,][]{Gehan2021}:
\begin{equation}
A^2(\nu_{m=0}) \propto \cos^2(i)
\end{equation}
\noindent and
\begin{equation}
A^2(\nu_{m=\pm 1}) \propto \frac{1}{2}\sin^2(i)\, .
\end{equation}
% Using the elements introduced in Sections~\ref{sec:mmf}, \ref{sec:zeta}, and \ref{sec:h_and_sigma} we produce a synthetic oscillation-mode PSD of a typical RGB star with $\nu_{\rm max}=108.3\, \mu{\rm Hz}$, $\Delta\nu=10.84\, \mu$Hz, and $\Delta\Pi_1=75.2$ seconds. The result is represented on the top panel of Fig.~\ref{fig:PSDs} by considering the typical observable frequency range [~$\nu_{\rm max}-3\Delta\nu$~:~$\nu_{\rm max}+3\Delta\nu$~].%, along with the $\zeta$ function.

\section{Calculation of the unperturbed mixed-mode frequency from magnetized PSD}
\label{App:nu_mm}

We rewrite the following expression:
\begin{equation}
    \nu^4-\nu_{{\rm obs}_0}\nu^3+\frac{\zeta A {\rm B}_0^2}{2}=0\, .
\end{equation}
\noindent as
\begin{equation}
        ax^4+bx^3+e=0\, ,
\end{equation}
by defining $a=1$, $b=-\nu_{{\rm obs}+0}$, and $e=\frac{\displaystyle \zeta}{\displaystyle 2}A{\rm B}^2_0$.

\noindent The general solution of this $4^{\rm th}$ order equation can be written as

\begin{equation}
    \nu = f\left(\nu_{{\rm obs}_0}\right)=-\frac{b}{4a} +\frac{p_4}{2} + \frac{\sqrt{p_5+p_6}}{2}\, ,
\end{equation}
\noindent with
$$p_1=27b^2e,$$
$$p_2=p1+\sqrt{-4(12ae)^3+p_1^2},$$
$$p_3=\frac{12ae}{\displaystyle3a\left(\frac{p_2}{2}\right)^{1/3}}+\frac{\displaystyle\left(\frac{p_2}{2}\right)^{1/3}}{3a},$$
$$p_4=\sqrt{\frac{b^2}{4a^2}+p_3},$$
$$p_5=\frac{b^2}{2a^2}-p_3,$$ 
$$p_6=-\frac{\displaystyle \frac{b^3}{a^3}}{4p_4}.$$

% aprox_numm1 = -b/(4*a) -p4/2 - np.sqrt(p5-p6)/2
% aprox_numm2 = -b/(4*a) -p4/2 + np.sqrt(p5-p6)/2
% aprox_numm3 = -b/(4*a) +p4/2 - np.sqrt(p5+p6)/2
% aprox_numm4 = -b/(4*a) +p4/2 + np.sqrt(p5+p6)/2
\section{Skewness of the dominant peak in the ${\rm d}\tau_{{\rm obs}_0}$-LS periodogram}
\label{app:skew}
\subsection{Skewness measurement}
By expressing $p_{\rm peak}$, the array of length N periods corresponding to the peak with maximum amplitude in the LS periodogram and $A_{\rm peak}$ the corresponding amplitude in the LS periodogram, we can measure the Fisher-Pearson coefficient of skewness of the peak distribution as a function of its mean value $\mu$ and standard deviation $\sigma$ through:
\begin{equation}
    S = \frac{\displaystyle \frac{1}{N}\left(\sum_{n=1}^N A_{{\rm peak}_n}(p_{{\rm peak}_n}-\mu)^3\right)}{\sigma^3}\, .
    \label{eq:skew}
\end{equation}%On the bottom panel the new p
For the ${\rm d}\tau_{\rm obs}$-stretched PSD of the simulated star represented in the first panel of Fig.~\ref{fig:LSmag0}, the $S$ value is about $0.39$. When using the appropriate ${\rm d}\pi_0$-stretching function via Eq.~\ref{eq:dpi0} the skewness is brought closer to zero (quasi-normal distribution), as $S\approx0.07$ in the middle panel of Fig.~~\ref{fig:LSmag0}.

% Current interpretations of the oscillation spectra have ignored to search for skewness in the LS periodogram maxima. We prove here that non-zero skewness can be used as a detection method for internal magnetism. This aspect has never been investigated during the process of measuring internal rotation rates, and we might therefore have missed magnetic fields signatures in already well-studied stars in \cite{Vrard2016} and \cite{Gehan2018}. 
% If an asymmetric signature in the dominant peak in the $\tau_{{\rm obs}_0}$-LS periodogram does now allow neither a direct measure of the magnetic shift of mixed-mode frequencies, nor a direct correction of the value of $\Delta\Pi_1$, it might be decisive in the process of flagging stars of potential interest for the search of internal magnetism. 

\subsection{Lombscargle periodograms}
% \label{app:LS_skew}
{We use a Lombscargle periodogram instead of the classical periodogram based on Fourier transforms because the stretched spectrum is unevenly spaced in the $\tau_{\rm obs}$ variable. The Lombscargle periodogram is a non-linear operator. We therefore ensure that the skewness of the dominant peak in the $\tau_{\rm obs}$-LS value measured for the magnetic model is indeed related to magnetism and not to a property of the LS periodogram. To do so, we perform a nearest neighbor resampling on the $\tau_0$-stretched PSD, by resampling the stretched PSD to the minimum period resolution in the stretched PSD. We then perform a periodogram of the resampled stretched PSD with the \texttt{scipy.signal.periodogram} Python function. First, we confirm that the measurement of $\Delta\Pi_1$ from this periodogram is compatible with the measurement from the $\tau_{\rm obs}$-LS periodogram, which confirms that using LS periodograms does not bias the measurement of $\Delta\Pi_1$. Second, the skewness of the dominant peak in the $\tau_{\rm obs}$ periodogram for the non-magnetized and magnetized models is respectively 0.02 and 0.45, which confirms that the skewness detected in the magnetized model in Section~\ref{sec:skew} is indeed due to magnetism and is independent of the use of the LS method.}

\section{${\rm d}\pi_1$-stretched PSD}
\label{App:pi_1}
We could try to do the analogous study carried out from Sec.~\ref{sec:probeB} with $m=0$ modes by focusing on $m=\pm 1$ oscillation modes instead. If the star is magnetized but not rotating, $m=\pm 1$ modes overlap and the stretched period $\pi_1$ can be defined as:

\begin{equation}
        {\rm d}\pi_1 = -\frac{1}{\zeta}\frac{{\rm d}\nu_{{\rm obs}}}{f_1(\nu_{{\rm obs}})^2} \frac{1}{\displaystyle1 - {3\zeta} \frac{A{\rm B}_0^2}{f_1(\nu_{{\rm obs}})^4} } \, ,
    \label{eq:dpi1}
\end{equation}
\noindent with $f_1(\nu_{{\rm obs}})$ as the solution for: 
\begin{equation}
        \nu^4-\nu_{{\rm obs}}\nu^3+{\zeta A {\rm B}_0^2}=0\, .
\end{equation}
However, if the star rotates, $f_1$ has to be rewritten as $f_{\pm 1}$ as it depends on the value of $m$, as the solution for:
\begin{equation}
        \nu^4-\nu_{{\rm obs}_{\pm1}}\nu^3+{\zeta A {\rm B}_0^2} \pm \zeta\delta\nu_{\rm rot}=0\, .
\end{equation}

When using $\pi_1$ instead of $\pi_0$, we must include the effect of rotation as well in our calculation of $\Delta\Pi_1$ when looking at $m=\pm1$ modes if the star is magnetized and shows rotational signatures. As rotation signature are very likely to emerge in most observed stars, we suggest that we always ought to base the measurement of $\Delta\Pi_1$ inside magnetized stars on $m=0$ modes in the PSD.%To avoid this situation, we focus on $m-0$ modes%These modes are also more difficult to study as the energy is split between the two in two, resulting in much lower amplitude in the LS of each of the $\pi_{1,\pm}$ stretched PSDs.

\section{Bayesian fitting}
\label{app:fit}

\begin{figure}[h]
    \centering
    \includegraphics[width=0.5\textwidth]{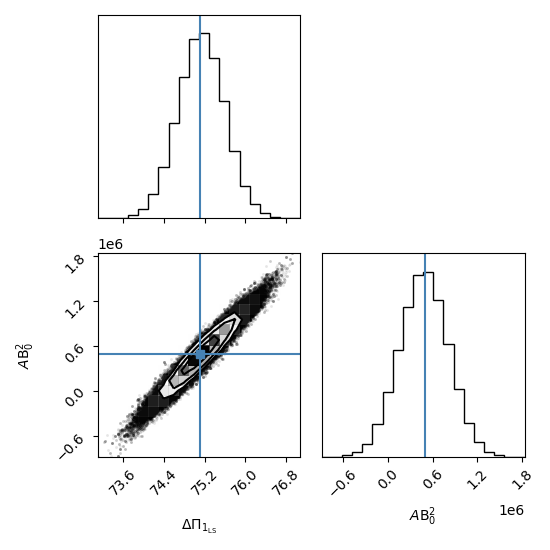}
    \caption{Corner plot showing the posterior distribution of $\Delta\Pi_1$ and $A{\rm B_0^2}$ obtained from the magnetized Bayesian fit (Eq.~\ref{eq:function}). Blue lines indicate the starting values from the $\chi_2^2$ fit.}
    \label{fig:fit_mag}
\end{figure}

\begin{figure}[h]
    \centering
    \includegraphics[width=0.25\textwidth]{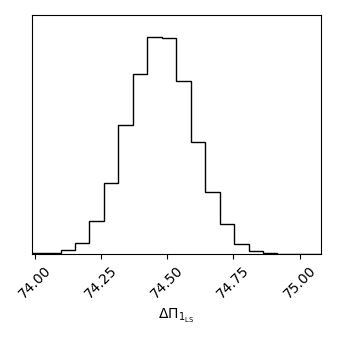}
    \caption{Corner plot showing the posterior distribution of $\Delta\Pi_1$ obtained from the non-magnetized Bayesian fit (Eq.~\ref{eq:function_nomag}). }
    \label{fig:fit_nomag}
\end{figure}

This appendix presents the corner plot of the distribution of the $\Delta\Pi_1$ and $A{\rm B}^2_0$ parameters ($\mathcal{F}$ model from Eq.~\ref{eq:function}, see Fig.~\ref{fig:fit_mag}) and $\Delta\Pi_1$ only ($\mathcal{F}_0$ nested model from Eq.~\ref{eq:function_nomag}, see Fig.~\ref{fig:fit_nomag}) from the Bayesian fitting process with the \texttt{emcee} python ensemble \citep{ForemanMackey2019}.

\section{Case of an inclined axisymmetric magnetic field}
\label{App:inclined}
Our study is focused on the detectability of the magnetized signature defined by Eq.~\ref{eq:Mathis}. This formulation approximates the effect of a stable, axisymmetric fossil field aligned with the rotation axis of the star, buried inside the radiate interior. However, spectropolarimetric observations of the surface of intermediate-mass stars on the MS show large-scale fossil magnetic fields inclined with {respect to} the rotation axis of the star \citep{Landstreet2000, Shultz2019}. As similar topologies are observed for stable fields at the surface of white dwarfs, it is necessary to extend the formalism to inclined stable magnetic fields. \cite{Loi2021} investigates the effect of such an inclined field on the mixed-mode frequencies of RGs. This study unveils, in good agreement with previous theoretical works such as that of \cite{Goode1992}, that an inclined magnetic field lifts the degeneracy of the azimuthal $m$ order and generates a hyperfine structure in the frequency spectrum, made up of nine components in the case of dipolar modes. Such a signature might seem very different from the asymmetric triplet studied here. However, the relative amplitudes of the peaks constituting the hyperfine structure still result in an asymmetric-dominated pattern when the field is inclined and has moderate amplitude \citep{Loi2021}. We therefore conclude that as long as the magnetic-field amplitude remains moderate \citep[as expected from the conservation of the magnetic flux from pre-main and main sequence dynamo actions; see][for more details]{Bugnet2021}, the method presented here can be applied even if the field is inclined inside the star.

\end{appendix}
\end{document}